\documentclass[12pt]{article}

\usepackage{amsmath}
\usepackage{titlesec}
\usepackage{setspace}
\usepackage{bm}
\usepackage{tikz}
\usepackage{scalerel}
\usepackage{graphicx}
\usepackage[labelfont=bf,labelsep=period]{caption}
\usepackage{soul}
\usepackage{natbib}

\usetikzlibrary{arrows,decorations.pathmorphing,backgrounds,positioning,fit}

\title{The Bayesian Synthetic Control: \\ Improved Counterfactual Estimation in the Social Sciences through Probabilistic Modeling}
\author{Elias Tuomaala*}

\begin{document}

\maketitle
\thispagestyle{empty}

\vspace*{\fill} \noindent *Elias Tuomaala holds a BA degree in Applied Mathematics from Harvard College, Cambridge, MA 02138
(elias.tuomaala@protonmail.com; ORCID: 0000-0003-0313-3960).
The research contained in this article was created as the author's undergraduate senior thesis, under the advising of
Dr. Rahul Dave from Harvard John A. Paulson School of Engineering.
The author would also like to thank Matthew Blackwell,
Associate Professor of Government at Harvard University,
for his invaluable feedback.
The author received no external funding,
and he has no financial interests or conflicts of interest to disclose.

\section*{Abstract}

\setcounter{page}{1}

Social scientists often study how a policy reform impacted a single targeted country.
Increasingly, this is done with the synthetic control method (SCM).
SCM models the country's counterfactual (non-reform or untreated) trajectory
as a weighted average of other countries' outcomes.
The method struggles to quantify uncertainty;
eg. it cannot produce confidence intervals.
It is also suspect to overfit.
We propose an alternative method, the Bayesian synthetic control
(BSC), which lacks these flaws.
Using MCMC sampling, we implement the method for two previously studied datasets.
The proposed method outperforms SCM in a simple test of predictive accuracy
and casts some doubt on significance of prior findings.
The studied reforms are the German reunification of 1990
and the California tobacco legislation of 1988.
BSC borrows its causal model,
the linear latent factor model, from the SCM literature.
Unlike SCM, BSC estimates the latent factors explicitly
through a dimensionality reduction.
All uncertainty is captured in the posterior distribution so that,
unlike for SCM, credible intervals are easily derived.
Further, BSC's reliability on the target panel dataset can be assessed
through a posterior predictive check;
SCM and its frequentist derivatives use up
the required information while testing statistical significance.

\vspace*{\fill} \noindent \textit{Key words}:
Treatment effect; Hierarchical model; Markov Chain Monte Carlo; German Re-unification; California Proposition 99

\newpage

\setcounter{tocdepth}{2}

\newpage

\section{Introduction}

\begin{figure}[h]
\centerline{\includegraphics[width=0.5\textwidth]{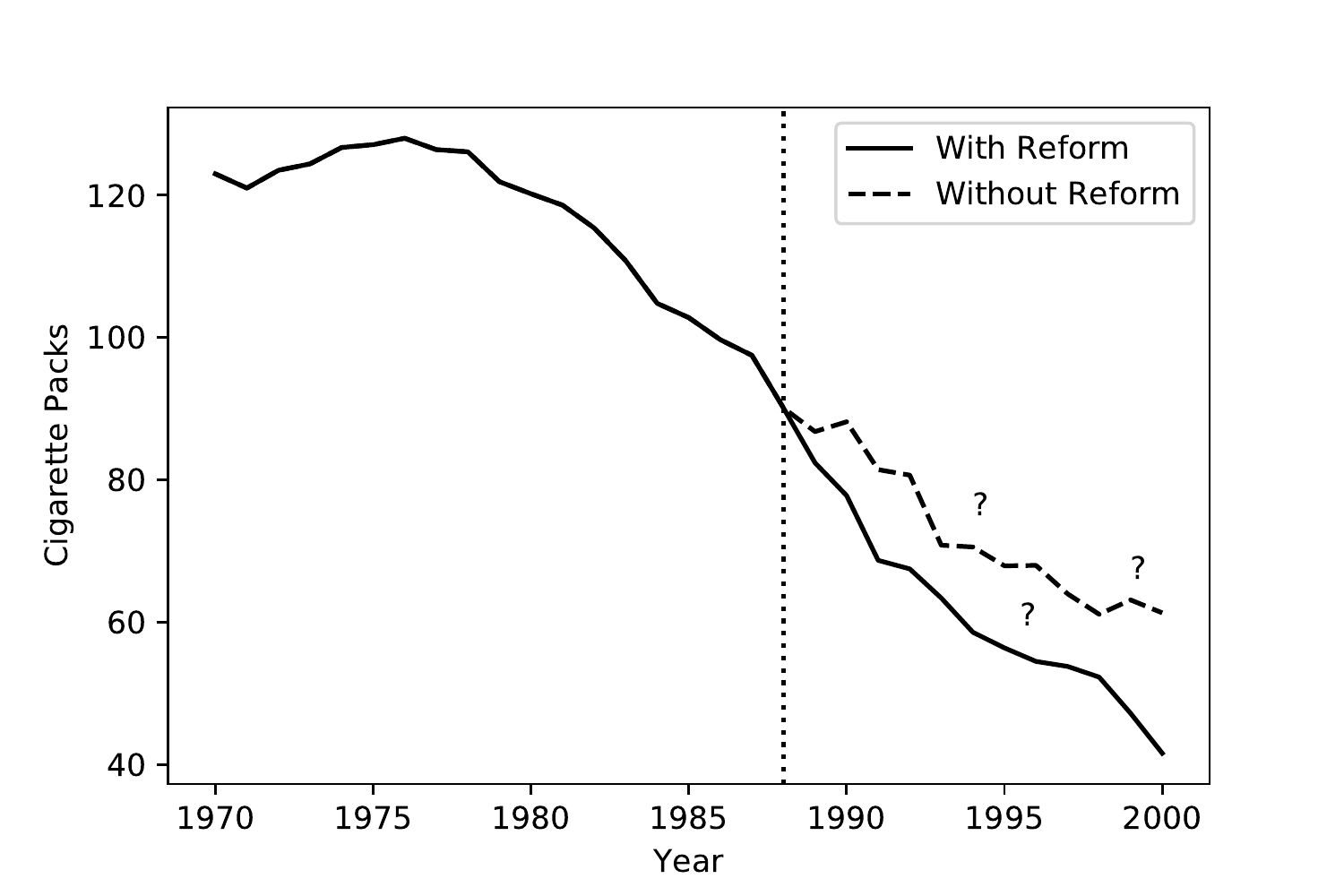}}
\caption{An example research setting for the synthetic control method (SCM):
the impact of California's 1988 tobacco control reform, Proposition 99.
The goal is to construct a synthetic counterfactual trajectory
to estimate what would have happened if the reform had never happened.}
\label{fig:motivatecali}
\end{figure}

Social scientists often need to study the lasting impact
of some one-off policy reform.
California's cigarette tax hikes of 1988
(visualized in Figure \ref{fig:motivatecali}) offer a concrete example:
how much was the state's smoking rate in the years since altered by the reform?
The question is challenging because we only observe one targeted state (a single treated unit),
so traditional econometric tools are inapplicable.
Researchers instead routinely apply a more recent technique,
the \textit{synthetic control method} (SCM)
\citep[see][]{abadie03,abadie10}.
The method constructs a synthetic version of the treated unit
as a weighted average of other, untreated, comparison units.
The weights are chosen such that the synthetic and observed country
resemble each other closely pre-reform.
The synthetic trajectory is then taken to estimate
the target unit's untreated trajectory post-reform, too.

SCM is popular and easy to use,
and in certain conditions its estimates are known to be asymptotically unbiased.
However, the tool has some important flaws.
Most notably, it struggles to quantify its own uncertainty.
Though it is associated with a simple significance test,
the tool cannot generate error bars or a confidence interval.
Second, SCM is suspect to overfitting concerns.
Data is noisy, both for the comparison units and for the target.
There will inevitably, then, be some weighted averages
that offer spurious though precise pre-reform fits.
SCM's frequentist nature leaves it vulnerable
to distortion due to such spurious correlations.

In this paper, we propose a novel Bayesian approach for constructing synthetic controls.
Our proposed method, the Bayesian Synthetic Control (BSC),
borrows SCM's motivating causal model
and then treats the estimation problem as a probablistic modeling exercise.
All estimate uncertainty is captured in the posterior distribution so that,
unlike SCM, the method readily produces any desired credible (confidence) intervals.
Point estimates are calculated by averaging over all plausible parameter values,
an approach that protects most standard Bayesian methods from the threat of overfitting
\cite[p.~147]{bishop06}.
The causal model, which is shared by SCM, its frequentist derivatives, and BSC,
supposes that all countries' trajectories are driven by a few
unobserved common trends.
Statistically, this corresponds to a latent linear factor model.
Unlike SCM, BSC estimates the latent factors explicitly.

We implement the proposed method for two previously studied research questions
using the original datasets.
First, we study the economic impact of German reunification in 1990 previously studied by
\citet{abadie15},
and exhibit that BSC outperforms SCM in a simple test of predictive accuracy.
In this application we also show how BSC can assess the validity of its modeling assumptions
through \textit{posterior predictive checking},
something SCM and its frequentist derivatives cannot do.
Then we examine the previously mentioned question of California tobacco controls of 1988
studied by \citet{abadie10} and \citet{benmichael18} among others.
In the latter application, we cast partial doubt on prior findings of statistical significance,
and demonstrate a method of endogenously selecting the number of latent factors included in the model.
We estimate the models computationally using standard
Markov Chain Monte Carlo (MCMC) posterior sampling.

The rest of this paper is structured as follows.
Section \ref{sec:scm} introduces the frequentist Synthetic Control Method
and reviews related literature.
In section \ref{sec:formal}, we introduce the full Bayesian Synthetic Control framework
and specify its underlying probabilistic latent variable model.
Sections \ref{sec:germany} and \ref{sec:california}
contain findings from the two empirical applications.
Section \ref{sec:conclusion} offers concluding remarks.

\section{Synthetic Control Method and Related Work}
\label{sec:scm}

\subsection{The Synthetic Control Estimator}

The SCM estimator was developed by \citet{abadie03}
and formalized further by \citet{abadie10}.
The estimator is a simple weighted average of
post-reform outcomes in the untreated comparison societies.
The control weights are derived by minimizing a prediction loss
on the pre-reform data.
The loss is calculated as a weighted sum of squared errors
in the variable of interest
and a number of separately selected covariates.
These covariates are arbitrarily chosen by the analyst
but should relate to the variable of interest.
They are included in the analysis for additional robustness
and may mitigate the threat of overfitting.

Formally, borrowing notation from \cite{abadie03},
consider a set of $J+1$ societies over $T$ years
such that society $1$ faces the treatment effect
in all years following some intermediate year $T_0$.
Denote by $y_{it}$ the outcome in society $i$ at time $t$
and by $\bm{z}_{i}$ a vector of $R$ society-specific, time-invariant covariates.
Collect values $y_{1t}$ for all $t \leq T_0$ and the entries of $\bm{z}_1$
into the vector $\bm{x}_1$ of length $R + T_0$.
Similarly, collect into the $(R + T_0) \times J$ matrix $\bm{X}_0$
the corresponding values for all other societies.
Let $\bm{V}$ be a matrix of variable weights for squared error calculation
(chosen separately, usually via cross-validation).
Finally, let $\bm{w}$ be the vector of $J$ comparison society weights.
Then, define the pre-intervention prediction loss as:

\begin{align}
||\bm{x}_1 - \bm{X}_0 \bm{w}||_{\bm{V}} = \sqrt{(\bm{x}_1 - \bm{X}_0 \bm{w})' \bm{V} (\bm{x}_1 - \bm{X}_0 \bm{w})}.
\end{align}

Then select the loss-minimizing synthetic control weights $\bm{w}^*$:

\begin{align}
\bm{w}^* &= \underset{\bm{w}}{\text{argmin}} ||\bm{x}_1 - \bm{X}_0 \bm{w}||_{\bm{V}}\\
&\text{such that} \\
w_j &\geq 0 \ \forall j \text{ and }\sum_{j=2}^{J + 1}{w_j} = 1.
\end{align}

Finally, use the weights $\bm{w}^*$ to calculate the SCM estimator $\hat{y}_{1t}^N$
for the counterfactual target society value in year $t$:
$\hat{y}_{1t}^N = \sum_{j=1}^{J+1} w_{j} y_{jt}$.

\citet{abadie10} show that this method can be motivated by a causal model.
They assume that a number of unobserved inter-society trends
drives change in the untreated outcome across all units.
Formally, if we denote the untreated outcome of society $i$
in year $t$ by $y^N_{it}$, the authors suppose that

\begin{align}
\label{eq:scmmotivate}
y^N_{it} &= \delta_t + \bm{\theta}_t \bm{z}_i + \bm{\lambda}_t \bm{\mu}_i + \varepsilon_{it}.
\end{align}

Here $\delta_t$ indicates an annual fixed effect,
$\bm{\theta}_t$ is a year-specific vector of coefficients,
$\bm{z}_i$ represents society-specific observable covariates,
$\bm{\lambda}_t$ is a year-specific vector of latent factors,
$\bm{\mu}_i$ denotes a vector of factor loadings for society $i$,
and $\varepsilon_{it}$ is an error term drawn from a distribution centered at zero.
This specification corresponds to a latent linear factor model
with unique factor loadings for each society.
Some of the loadings are fixed to match the society's
time-invariant covariate values.
Though SCM doesn't estimate any model parameters explicitly,
the authors use the model to prove that
the SCM estimator is asymptotically unbiased
under certain assumptions.

\begin{figure}
\centerline{\includegraphics[width=0.5\textwidth]{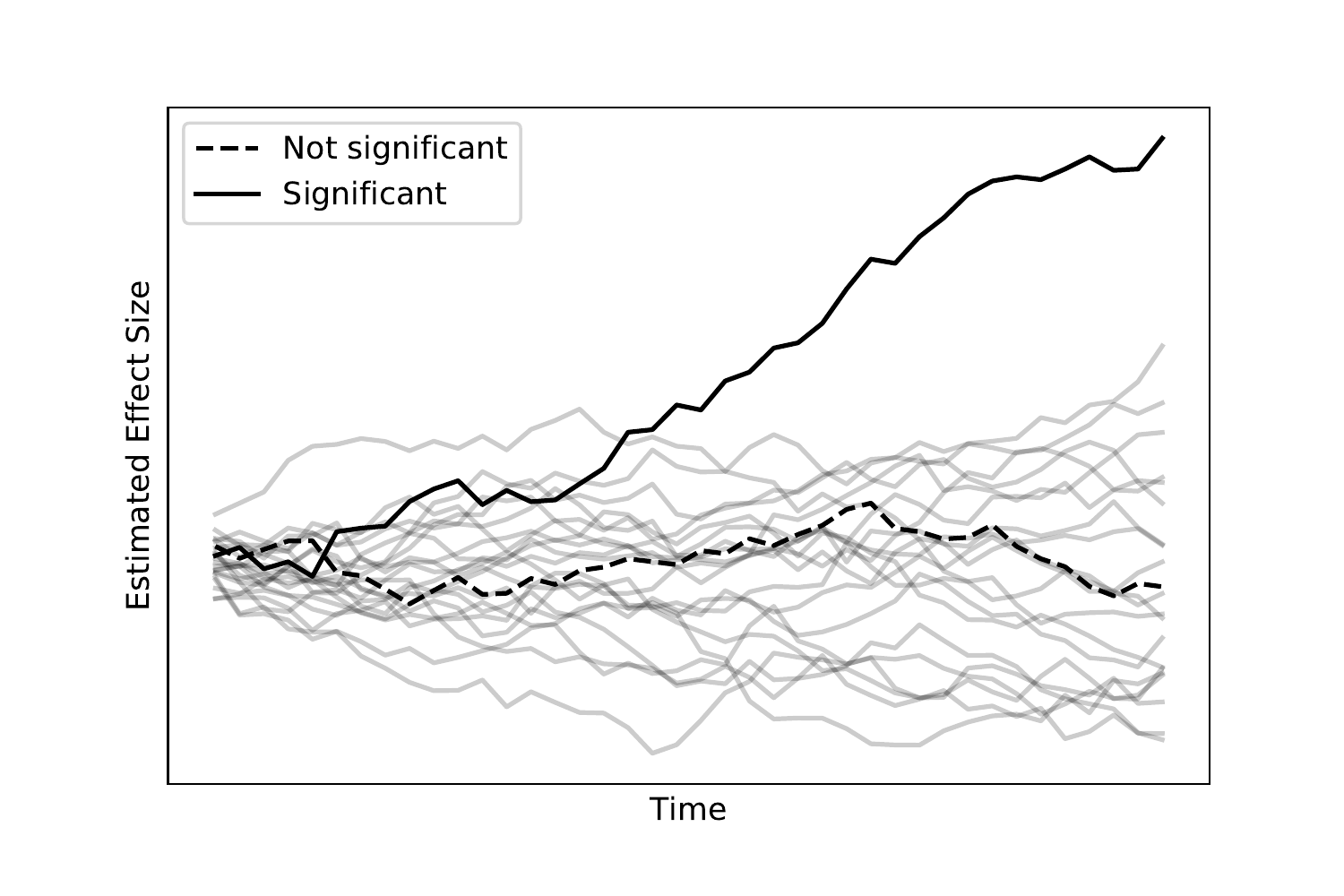}}
\caption{SCM significance test.
The light lines are post-reform prediction errors for comparison societies (placebo errors).
The dark ones are two alternative post-reform treatment effect estimates for the target society.
The solid is considered significant, the dashed is not.}
\label{fig:scmsignif}
\end{figure}

\citet{abadie10} also propose a relabeling-based signficance test for SCM.
It involves constructing an SCM estimate for each of the comparison societies.
Findings are considered statistically significant if
the estimated effect is larger for the target society than for 95\% of the others.
Figure \ref{fig:scmsignif} illustrates this relabeling test.

\subsection{Extensions and Related Work}

In addition to numerous applied studies that use SCM
\citep[eg.][]{aytug17,barlow18,karlsson15},
recent work has further studied and extended the methodology itself.
\citet{xu17} proposes the Generalized Synthetic Control (GSC),
which constructs the synthetic counterfactual
only after deriving an explicit point estimate for the latent factors.
\cite{ferman16} show that SCM becomes asymptotically biased
if the pre-treatment match is not exact.
In response, \cite{benmichael18} propose an extension to SCM,
the Augmented Synthetic Control Method (ASCM),
which includes a bias-reducing (though not eliminating) correction term.

Separately, \cite{hsiao12} develop a related panel data approach (PDA).
They set up a generic latent linear factor model
and show that the SCM model is its special case.
Unlike SCM, PDA considers no covariate features.
The authors find simulation-based evidence of worsening overfit
as the number of comparison societies grows.
For estimation, they use regression instead of weighted average optimization.
\cite{abadie15} demonstrate that regression-based methods
can unnecessarily rely on extrapolation when interpolation would suffice.
Yet, \citet{wan18} show that PDA is more robust to certain assumption violations than SCM.

GSC bootstraps a confidence interval and ASCM yields a standard error,
and PDA calculates a t-statistic based on estimated predictor variance.
All are asymptotically valid,
but none of the authors derive their predictor's full distribution.
The asymptotic behavior may thus fail if the pre-treatment period is not very long.
Further, GSC and ASCM base their confidence intervals
on the errors generated for comparison societies when relabeling.
This prematurely uses up information that could otherwise have been used
to check the method's applicability to a particular dataset.
(This task is made important by the strong assumption
that a linear factor model suffices to describe the data generating process.)
An obvious way to do the check is to see whether the comparison society outcomes
generally fall within their predicted confidence intervals while relabeling.
This cannot be done for GSC or ASCM,
because comparison society confidence intervals are by design inflated to be wide enough to include the observation.
Further, none of GSC, ASCM, or PDA proposes an approach
to eliminate the threat of overfitting.

Finally, a research team at Google has made an early
Bayesian contribution that relates to the synthetic control literature
\citep{brodersen15}.
In their paper, the authors describe a very general
Bayesian state-space model for cross-sectional time-series variables.
Inspired by synthetic controls, the authors include a regression on comparison units.
The tool doesn't have the flaws of the frequentist SCM derivatives
(lack of confidence intervals, overfit, inability to check model applicability).
Their approach is designed for advertising research, however,
and lacks a causal model applicable to study of policy reforms.

\section{The Bayesian Synthetic Control}
\label{sec:formal}

\subsection{Causal Model}
\label{subsec:notation}

Consider a set of $T$ years and $J$ societies
and some quantitatively measured outcome of interest.
Suppose that the outcome in certain years and societies was impacted
by the unknown treatment effect of some policy intervention.
Denote by $y_{it}^T$ and $y_{it}^N$ the outcome in society $i$ and year $t$
in the presence and absence of the treatment, respectively,
and by $y_{it}$ the actual observed outcome.
Suppose that change over time in the untreated outcome is driven
by some $L$ latent inter-society trends (where $L < J$) in a linear fashion.
We can then write that

\begin{align}
\label{eq:basis_general}
y_{it}^N &= \delta_{t} + \kappa_{i} + \bm{f}_{t}' \bm{\beta}_{i} + \varepsilon_{it},\\
y_{it} &= y_{it}^N + \alpha_{it} d_{it}.
\end{align}

Here $\delta_{t}$ and $\kappa_{i}$ are year and society fixed effects;
$\bm{f}_{t}$ captures the factor values for the year $t$;
$\bm{\beta}_{i}$ is the vector of time-invariant factor loadings of society $i$;
$\varepsilon_{it}$ is random noise;
$\alpha_{it}$ denotes the treatment effect in year $t$ and society $i$;
and $d_{it}$ is an indicator for whether society $t$ was treated in year $i$.

BSC's model is thus a special case of that of \citet{hsiao12}:
one of its factors is here forced to have constant unit loadings to create the year fixed effect.
This is done for consistency with the SCM model of \citet{abadie10}.
The BSC model differs from the latter on two points only.
BSC forces one factor to be constant over time to create a society fixed effect.
More notionally, BSC also withholds from identifying
a subset of the factor loadings with observed  society-specific covariates.

We can stack the individual parameters together into higher-dimensional
vectors and matrices:
let $\bm{B}$ denote the $J \times M$ transformation matrix of factor loadings;
let $\bm{F}$ denote the $T \times M$ matrix of factor values over time;
let $\bm{\Delta}$ and $\bm{K}$ denote the $T \times J$ year and society fixed effect matrices, respectively;
let $\bm{\varepsilon}$ denote the $T \times J$ matrix of random noise;
let $\bm{A}$ and $\bm{D}$ denote the $T \times J$ treatment effect and indicator matrices, respectively;
and let $\bm{Y}$ denote the $T \times J$ dimensional outcome matrix.
Use $\circ$ for the elementwise (Hadamard) matrix product.
Then we can describe the whole system thus:

\begin{align}
&\bm{Y}^N = \bm{\Delta} + \bm{K} + \bm{F} \bm{B}' + \bm{\varepsilon},\\
&\bm{Y} = \bm{Y}^N + \bm{A} \circ \bm{D}.
\end{align}

\subsection{Estimation Goal}
\label{subsec:probmodel}

We observe directly $\bm{Y}$
and assert $\bm{D}$ as a known model parameter.
Typically, a single society faces the treatment effect
for all years following some known start year $T_0$.
We want to investigate the untreated (reform-free) counterfactual,
i.e. the posterior predictive distribution $p(\bm{Y}^N | \bm{Y})$.
If we denote by $\bm{\theta}$ the collection of all model parameters
(and assume the noise term independent from them),
we can also express this as $\int p(\bm{Y}^N | \bm{\theta}) p(\bm{\theta} | \bm{Y}) d\bm{\theta} $
or, up to a normalizing constant, as
$\int p(\bm{Y}^N | \bm{\theta}) p(\bm{Y} | \bm{\theta}) p(\bm{\theta}) d\bm{\theta} $.

As soon as we set prior distributions for the parameters,
the above integral is well defined.
It is necessarily too complex for exact solving, though,
so in practice it must be approximated computationally.
The empirical sections of this paper do so using Markov Chain Monte Carlo (MCMC).
MCMC approximates a distribution
by sending a sampler on a random walk on the parameter space.
Though widely used and reliable,
MCMC struggles with multimodal distributions.
This flaw has an important implication on one of the priors
set in Section \ref{subsec:distros}.

\subsection{Distributional Assumptions}
\label{subsec:distros}

The prior distribution structure is a defining component of BSC.
It is visualized using directed graph notation in Figure \ref{fig:directed}
and recorded in detail in the Supplementary Material.
All random noise is assumed to follow a normal distribution with unknown constant variance.
Gaussian priors are used for most other parameters,
with the notable exception of the noise standard deviation
and hyperparameter standard deviations.
For them, the half-Cauchy distribution is used instead;
\citet{polson12} argue that half-Cauchy's fat tail
makes it the most appropriate prior for scaling parameters.

\begin{figure}
\centering
\begin{tikzpicture}
\tikzstyle{v}=[circle, minimum size=2mm,draw,thick]
\draw (-2, 0.5) rectangle (1.1, 2.5);
\node[inner sep=0pt] (legend-T) at (-1.12, 2) {$T$ \textit{years}};
\node[inner sep=0pt] (legend-J) at (-0.89, 1.5) {$J$ \textit{societies}};
\node[inner sep=0pt] (legend-M) at (-0.45, 1) {$L$ \textit{latent factors}};
\node[inner sep=0pt] (b-distro) at (-0.3, -4) {\includegraphics[width=.12\textwidth]{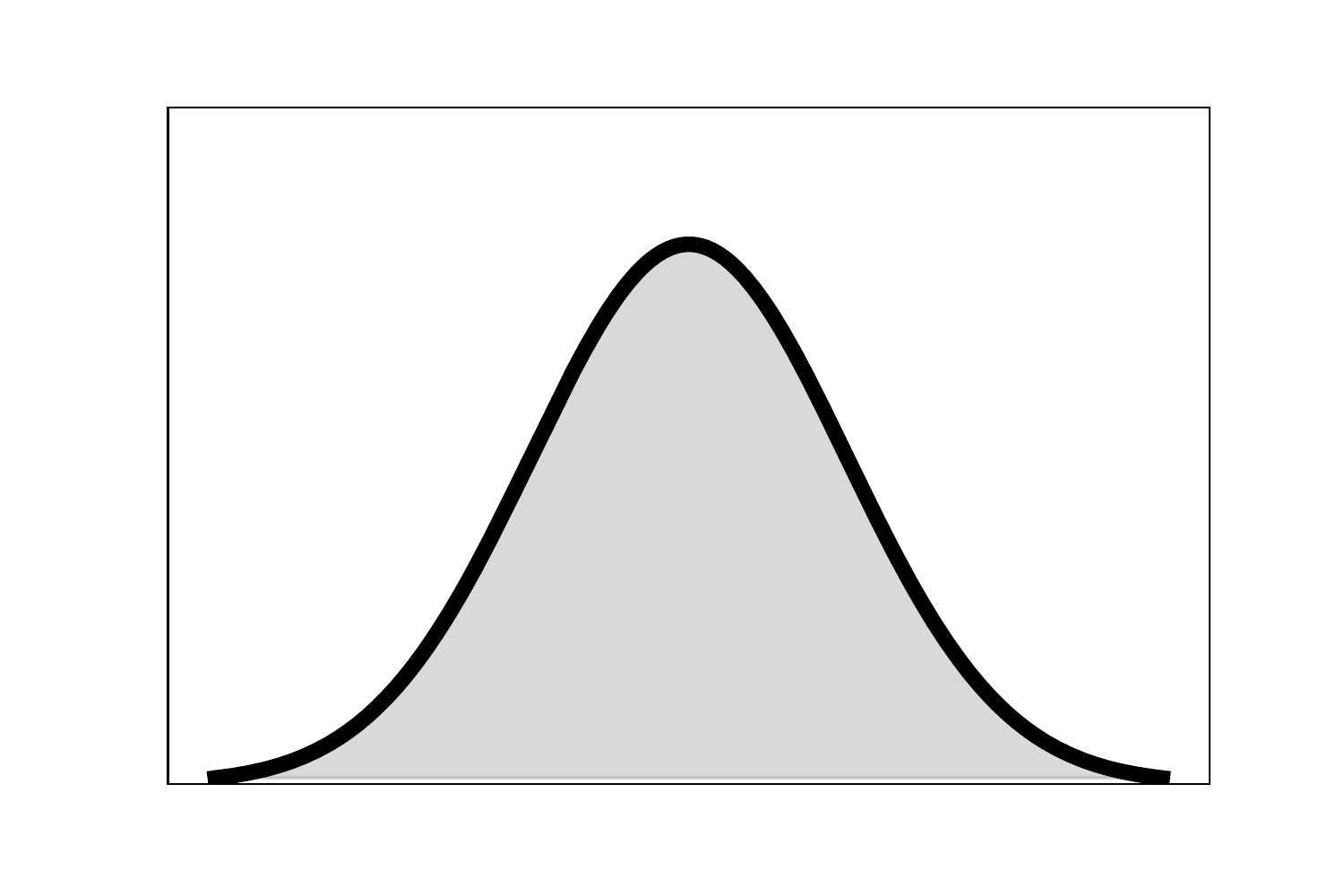}};
\node[inner sep=0pt] (d-distro) at (6.30, 2.2) {\includegraphics[width=.12\textwidth]{figure3_gauss.pdf}};
\node[inner sep=0pt] (F-distro) at (2.70, 2.2) {\includegraphics[width=.12\textwidth]{figure3_gauss.pdf}};
\node[inner sep=0pt] (a-distro) at (4.5, 2.2) {\includegraphics[width=.12\textwidth]{figure3_gauss.pdf}};
\node[inner sep=0pt] (k-distro) at (9.2, -4) {\includegraphics[width=.12\textwidth]{figure3_gauss.pdf}};
\node[inner sep=0pt] (s-distro) at (-0.3, -7) {\includegraphics[width=.12\textwidth]{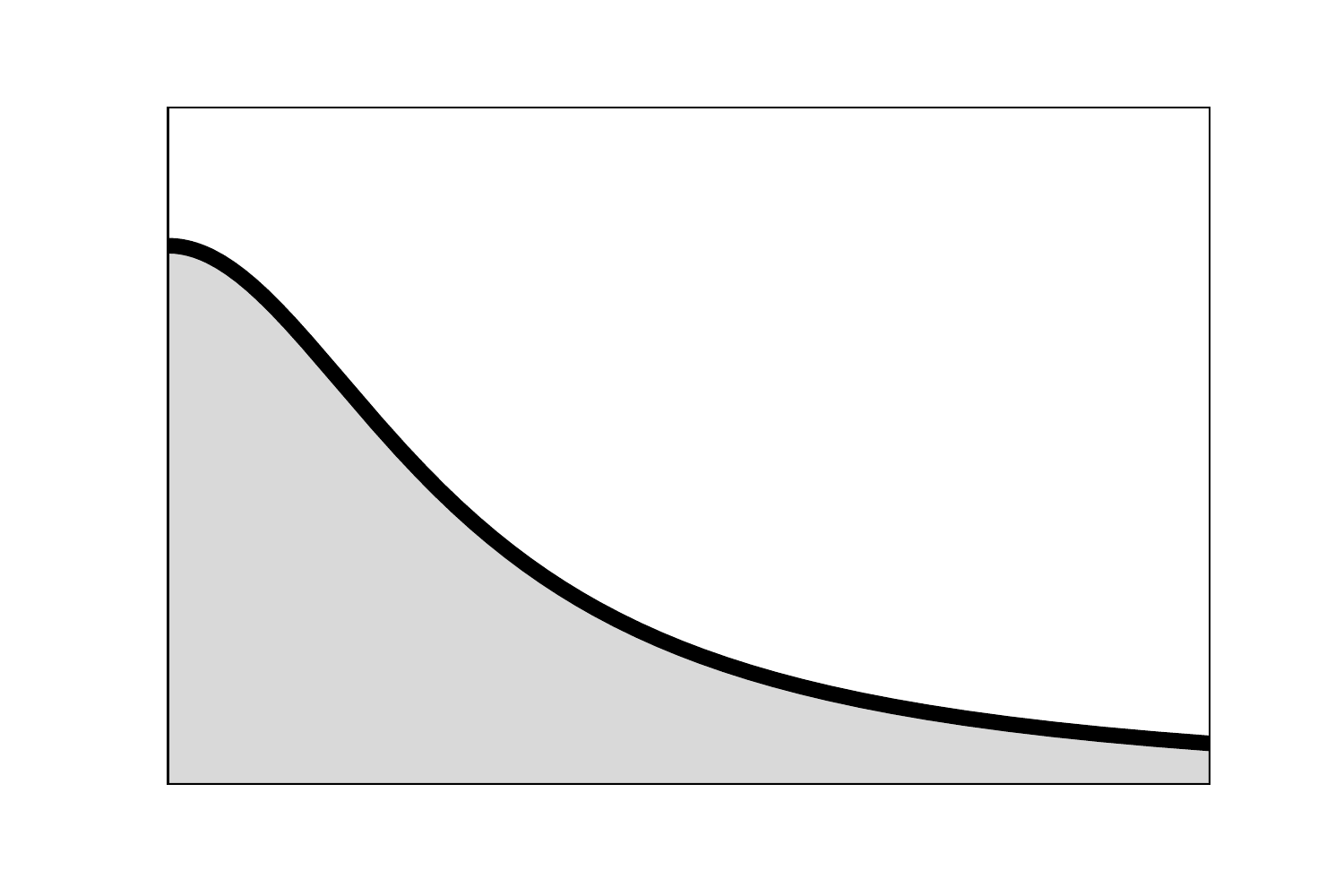}};
\node[inner sep=0pt] (y-distro) at (5.8, -10) {\includegraphics[width=.12\textwidth]{figure3_gauss.pdf}};
\node[inner sep=0pt] (bsd-distro) at (0.05, -0.6) {\includegraphics[width=.12\textwidth]{figure3_hcauchy.pdf}};
\node[inner sep=0pt] (ksd-distro) at (9.25, -0.6) {\includegraphics[width=.12\textwidth]{figure3_hcauchy.pdf}};
\node[inner sep=0pt] (bmu-distro) at (1.95, -0.6) {\includegraphics[width=.12\textwidth]{figure3_gauss.pdf}};
\node[inner sep=0pt] (kmu-distro) at (7.5, -0.6) {\includegraphics[width=.12\textwidth]{figure3_gauss.pdf}};
\node (F-dim) at (2.70, 1.55) {\scaleto{T \times L}{4pt}};
\node (d-dim) at (6.3, 1.55) {\scaleto{T \times 1}{4pt}};
\node (a-dim) at (4.5, 1.55) {\scaleto{T \times J}{4pt}};
\node (bsd-dim) at (0.05, -1.28) {\scaleto{L \times 1}{4pt}};
\node (ksd-dim) at (9.25, -1.28) {\scaleto{1 \times 1}{4pt}};
\node (bmu-dim) at (1.95, -1.28) {\scaleto{L \times 1}{4pt}};
\node (kmu-dim) at (7.5, -1.28) {\scaleto{1 \times 1}{4pt}};
\node (b-dim) at (-0.3, -4.7) {\scaleto{J \times L}{4pt}};
\node (k-dim) at (9.2, -4.7) {\scaleto{J \times 1}{4pt}};
\node (s-dim) at (-0.3, -7.7) {\scaleto{1 \times 1}{4pt}};
\node (y-dim) at (5.8, -10.7) {\scaleto{T \times J}{4pt}};
\node (mu-eq) at (7, -7) {\scaleto{= \bm{F} \bm{B}' + \bm{\Delta} + \bm{K} + \bm{A} \circ \bm{D}}{8pt}};
\node (y-eq) at (5.8, -9.2) {\scaleto{\sim \mathcal{N} \left( \bm{M}, \bm{\Sigma} \right)}{8pt}};
\node (ss-eq) at (3.55, -8.5) {\scaleto{= \sigma \bm{1_{[T \times J]}}}{11pt}};
\node (K-eq) at (7.7, -5.5) {\scaleto{= \bm{1_{[T]}} \otimes \bm{\kappa} }{11pt}};
\node (D-dim) at (6.7, -3) {\scaleto{= \bm{\delta} \otimes \bm{1_{[J]}} }{11pt}};
\node[v] (bmu) at (1.95,-2) {$\bm{\beta^{\mu}}$};
\node[v] (bsd) at (0.05,-2) {$\bm{\beta^{sd}}$};
\node[v] (kmu) at (7.5,-2) {$\kappa^{\mu}$};
\node[v] (ksd) at (9.25,-2) {$\kappa^{sd}$};
\node[v] (b) at (1,-4) {$\bm{B}$};
\node[v] (F) at (2.75,1) {$\bm{F}$};
\node[v] (d) at (6.25,1) {$\bm{\delta}$};
\node[v] (a) at (4.5,1) {$\bm{A}$};
\node[v] (D) at (5.375,-3) {$\bm{\Delta}$};
\node[v] (k) at (8,-4) {$\bm{\kappa}$};
\node[v] (K) at (6.25,-5.5) {$\bm{K}$};
\node[v] (m) at (4.5,-7) {$\bm{M}$};
\node[v] (s) at (1, -7) {$\sigma$};
\node[v] (ss) at (2.2, -8.5) {$\bm{\Sigma}$};
\node[v] (y) at (4.5, -10) {$\bm{Y}$};
\draw [->] (bmu) to (b);
\draw [->] (bsd) to (b);
\draw [->] (kmu) to (k);
\draw [->] (ksd) to (k);
\draw [->] (b) to (m);
\draw [->] (F) to (m);
\draw [->] (d) to (D);
\draw [->] (D) to (m);
\draw [->] (k) to (K);
\draw [->] (K) to (m);
\draw [->] (a) to (m);
\draw [->] (m) to (y);
\draw [->] (s) to (ss);
\draw [->] (ss) to (y);
\end{tikzpicture}
\caption{The BSC probabilistic model as a directed graph.
Nodes and arrows represent model variables and their conditional dependencies.
For non-deterministic variables, vector/matrix dimension is included
and the prior distribution (Normal/half-Cauchy, conditional on previous variables) visualized.
All Gaussians have elementwise variances instead of covariance matrices.
Outer and Hadamard products are denoted by $\otimes$ and $\circ$.
Boldface denotes non-scalar; capital case denotes a matrix;
$\bf{1}_{[A]} / \bf{1}_{[A \times B]}$ is a vector/matrix of ones.
The full model definition is included in the Supplementary Materials.}
\label{fig:directed}
\end{figure}

For the society fixed effects $\kappa_i$ and factor loadings $\beta_{im}$,
the prior is hierarchical.
This means that a single Gaussian prior is used for all societies' values,
but such that its mean and variance are estimated endogenously within the model.
Hierarchy biases the model against outlier values.
Importantly, this prior disfavors setting
anomalous factor loadings for the target society.
This amounts to a soft constraint against extrapolation,
a form of prediction which the developers of SCM \citep{abadie03,abadie10}
fear less reliable than interpolation.

Most priors should be uninformative or reflect outside information.
However, if the $L$ latent factors are left with identical uninformative priors,
they become rotationally nonidentifiable.
The nonidentifiability artifically gives the posterior distribution $L!$ identical modes,
a feature that massively slows down computation for MCMC sampling.
The implementations in this paper therefore
match each factor with a unique, strongly informative prior.
Each is a Gaussian centered around a different Principal Component Analysis (PCA) base vector.
The choice of shapes was made because PCA is known to be the maximum-likelihood estimate
of the latent linear factor model \citep[p. 147]{bishop06}.
The PCA components are derived, before MCMC sampling,
from the non-treated societies' data.
This informative prior solution has substantial flaws
(mostly, it prevents the model from exploring other plausible factor modes)
but is necessary barring a change from conventional MCMC sampling to another estimation strategy.

Finally, it is crucial that the treatment effect terms have near-uniform priors.
Otherwise the treated data would contaminate
the target society's factor loading estimates.
That would contradict the core synthetic control idea
of deriving a pattern from pre-reform data and applying it post-reform.

\section{Application: German Reunification}
\label{sec:germany}

\subsection{Background}

One of the important papers on the original SCM methodology
investigates the impact of the German reunification in 1990
\citep{abadie15}.
They study the effect on per capita incomes in former West Germany.
The reunification amounted to the West merging with a poorer country,
so the impact ought to have been negative.
Indeed, the authors find that by 2003,
West German per capita GDP would have been 12\% higher without the reform.
We implement the BSC framework to examine this same research question.

\subsection{Data}

We base our work on the dataset used by the original authors
which is released to the public domain
\citep{hainmueller14}.
The target variable is per capita GDP adjusted for purchasing power parity (PPP).
The data is originally acquired from the OECD National Accounts and
Germany's Federal Statistical Office.
The dataset also includes five other covariates useful for SCM,
though they play no part in the BSC implementation.
The data covers West Germany and 16 OECD countries:
all 23 member states from 1990,
barring seven which the authors excluded due to anomalous economic development.
For consistency with the prior work,
we use the same set of 17 countries.
The study covers years 1960-2003,
of which 1990-2003 are considered treated for West Germany.

We make one alteration to the dataset.
The original authors measure GDP in current US dollars
rather than ones adjusted for inflation.
Consequently, the variable grows at an arftifically high exponential rate.
This interacts poorly with the BSC assumption that
the random noise term's variance is constant over time.
One would expect the magnitude of the random error
to be more or less proportional to the outcome scale.
To moderate this issue, we adjust the GDP per capita figures approximately for inflation
and express them in constant 2003 US Dollars.
To do so, we use the US GDP deflator time series
recorded from The World Bank's World Development Indicators
\citep{wdi19}.
We rerun the replication code of \cite{abadie15}
on the inflation-adjusted dataset,
and base all BSC-SCM comparisons on the resulting SCM findings.
They remain unchanged up to the precision of previously published figures.

\subsection{Parameter Specification}

The most notable parameter specification questions relate to the latent factors.
For simplicity and ease of computation,
this application presumes their number to be small at $L = 4$.
The first four PCA components are able to explain 0.997 of all variance in data,
so the number appears sufficient.
The prior for each factor is given a standard deviation greater
by a factor of two than the standard deviation of the associated PCA component.
(An attempt to use a factor of three
failed to ensure factor identification in the posterior distribution.)
The full prior distribution specification for all parameters
can be found in the Supplementary Materials.

\subsection{Findings}

\begin{figure}
\centerline{\includegraphics[width=0.6\textwidth]{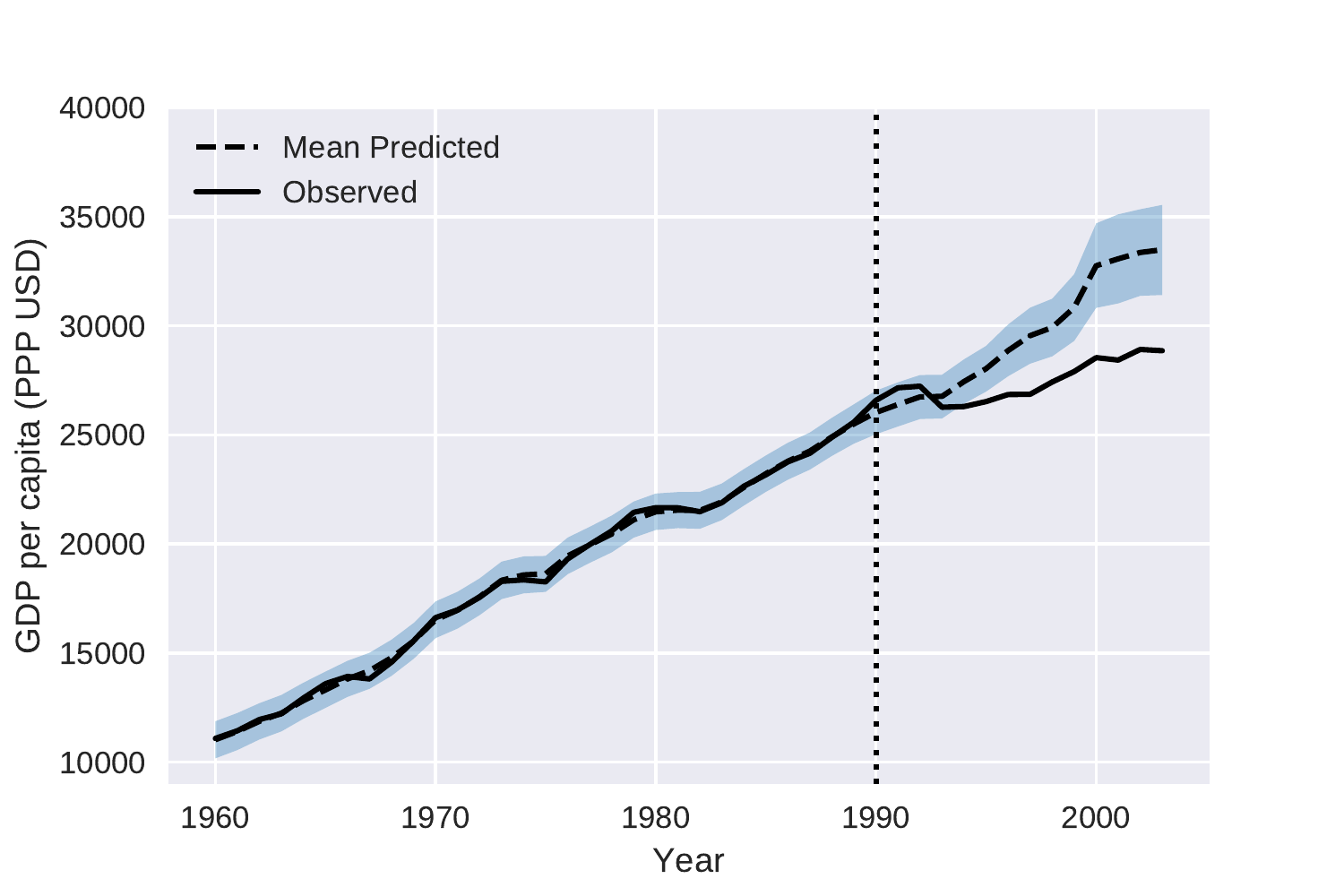}}
\caption{The BSC posterior predictive distribution of the counterfactual
West German per capita GDP (2003 PPP USD) without reunification.
The dashed trajectory and the shaded region represent the distribution's
mean and 95\%-confidence interval (CI).
The CI includes error re-sampling for the full time period.
Observed data is represented by the solid line.}
\label{fig:deubsccf}
\end{figure}

Figure \ref{fig:deubsccf} summarizes the empirical findings:
it visualizes the resulting BSC counterfactual estimate
(posterior predictive distribution) of the West German per capita GDP.
The findings are largely in line with those of
\cite{abadie15}.
We find that the counterfactual growth trajectory doesn't much
differ from the observed data in the first four years post-reform.
Starting in 1994, however, the two trajectories diverge.
By 2003, the observed GDP level is some USD 4,630,
or 16.0\%, below the mean predicted counterfactual value.
This is equivalent to a fall in the average annual growth rate by 1.1 percentage points,
from over 1.9\% to just under 0.9\%.
SCM predicts a similar though slightly smaller gap:
USD 3,360, or 11.7\%, which corresponds to a 0.7 fall in the average growth rate.

\begin{figure}
\centerline{\includegraphics[width=0.50\textwidth]{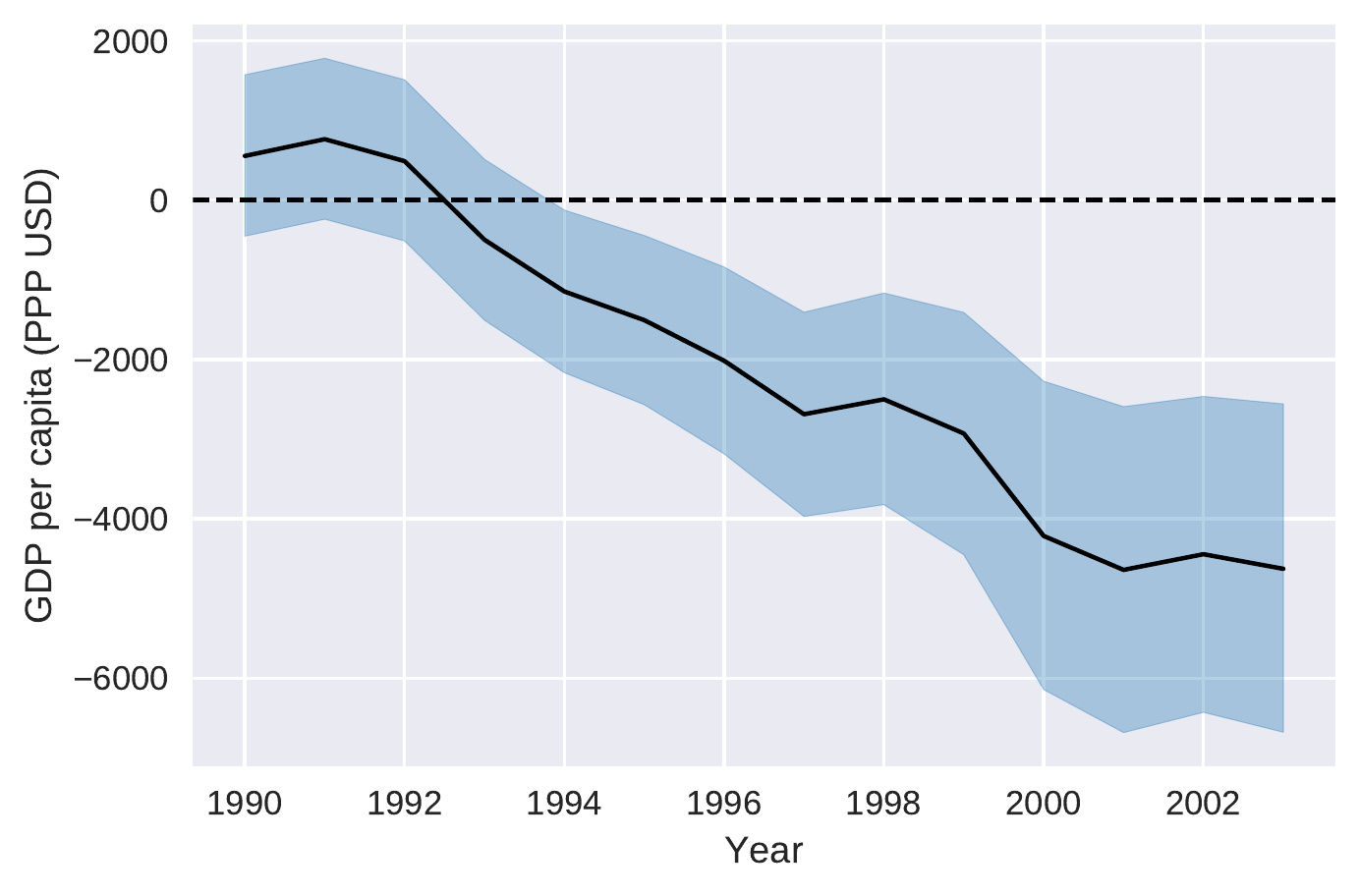}\includegraphics[width=0.50\textwidth]{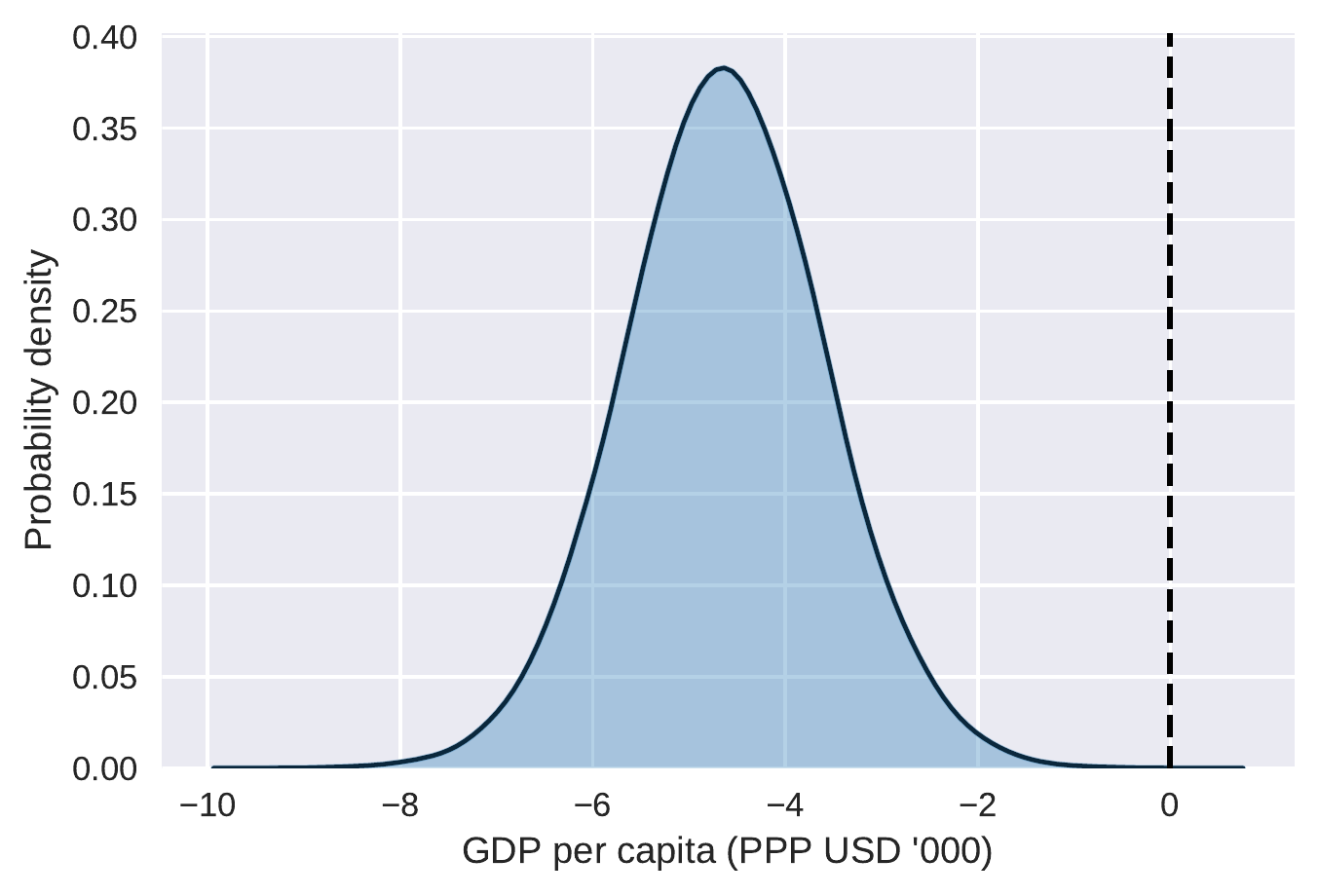}}
\caption{BSC posterior distribution of the reunification's treatment effect
on West German per capita GDP (2003 PPP USD).
Left panel plots the mean and 95\%-CI 
of the effect by year.
Right panel visualizes the full distribution
of the treatment effect at the year 2003,
i.e. the aggregate effect over the full post-reform time period.}
\label{fig:deualpha}
\end{figure}

Importantly, from 1994 onwards, the observed data falls far outside the credible
interval (CI) of the posterior predictive distribution.
By 2003, the 95\%-CI of the treatment effect  is USD 2,570 - 6,680.
Figure \ref{fig:deualpha} illustrates this in more detail.
On the left it depicts the treatment effect's mean and 95\%-CI by year;
the right panel draws the full posterior distribution for the effect by the end year 2003.
The figure illustrates how the treatment effect grew significant around 1994.
It also shows how BSC gives a near-zero probability for an overall nonnegative treatment effect.
The West German per capita income almost certainly fell due to reunification.

\begin{figure}
\centering
\includegraphics[width=0.6\textwidth]{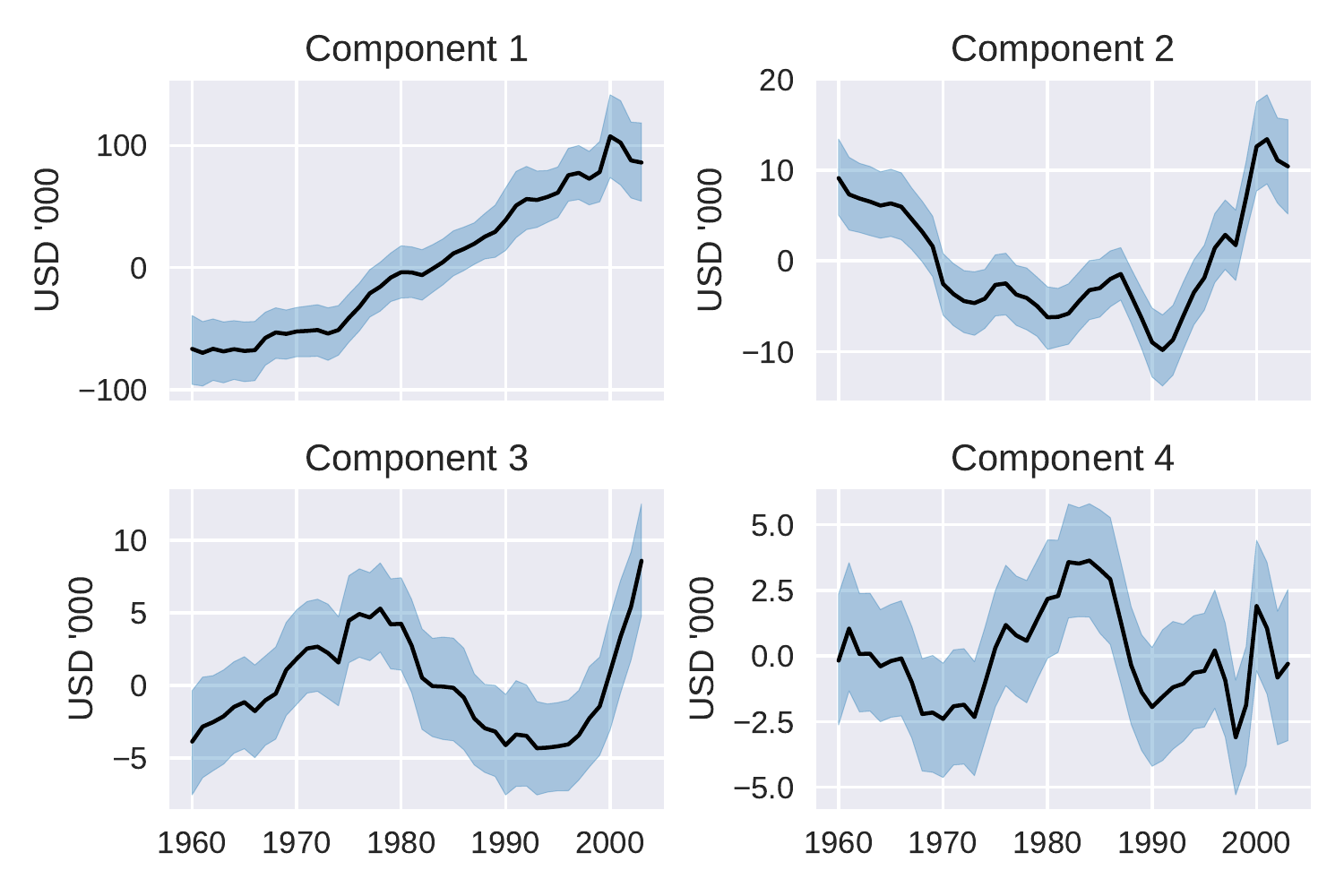}
\caption{BSC posterior distribution of each of the latent factors.
The solid trajectory and the shaded region
represent the mean and the 95\%-CI of the distribution.
The factors are numbered in the order of descending variance of the PCA component
around wich the prior of the factor is centered.}
\label{fig:deulatents}
\end{figure}

As a plausibly interesting side effect,
BSC also generates full probabilistic estimates of all other model parameters.
(In this it differs substantially from its frequntist counterparts.)
An example is provided in Figure \ref{fig:deulatents}
which graphs each of the latent factor posterior distributions.
Closer analysis could identify correspondence between
these factors and observed international trends,
such as global overall productivity growth (Component 1)
or energy prices (possibly Component 3).
Using the factor loading estimates, for their part,
we could measure similarity between countries and group them into clusters.

BSC's ability to yield findings on the other parameters could in this fashion
prove useful for political scientific attempts
to bridge the gap between quantitative findings and qualitative interpretation or explanation,
i.e. to "put qualitative flesh on quantitative bones" \citep{tarrow95}.
\cite{abadie15} think this one of the major goals of the synthetic control approach.
It must be noted, though, that these secondary parameter estimates
do not contain any additional information on the size of the treatment effect.

\subsection{Prediction Accuracy Comparison to SCM}

\begin{figure}
\centering
\includegraphics[width=0.6\textwidth]{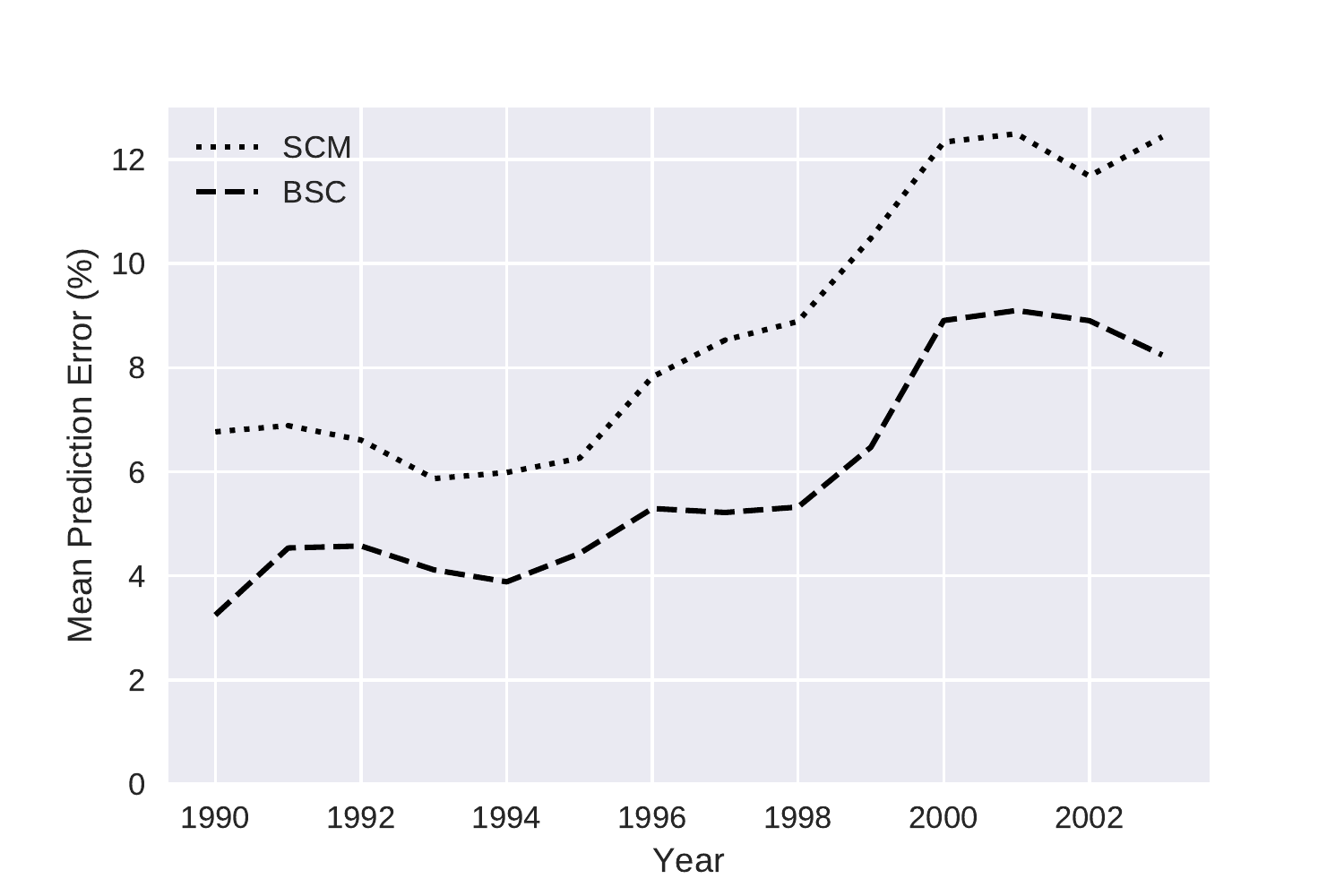}
\caption{Accuracy comparison of BSC and SCM.
The dotted line maps the average absolute error of SCM
when the method is used to predict the comparison societies'
outcomes in each year.
The dashed line does so for BSC.
Error is measured as percentage of the observed outcome value.}
\label{fig:deu_accuracy}
\end{figure}

A relabeling exercise provides an excellent opportunity
for accuracy comparison between BSC and SCM.
To do so, we use each framework to predict in turn
the observed post-treatment trajectory of each of the comparison societies.
At each run, we measure the distance between the prediction and the observation.
The more accurate the method, the shorter the distance ought to be.
To do so with BSC, we use the posterior predictive mean as our point estimate.
We visualize the comparison society average of the error for each year and framework in Figure \ref{fig:deu_accuracy}.
Overall, the findings show that BSC exhibits greater predictive accuracy on this dataset.
In most years, it is on average over two percentage points closer to reality than SCM.
A likely explanation for the improvement in accuracy
is BSC's freedom of overfit concerns.
Evidence on further datasets and simulations is required
for a more reliable comparison.

\subsection{Model Validity Checking}

A linear factor model makes strong assumptions
about the data-generating process,
assumptions which are bound to be more or less violated in reality.
The severity of those violations,
and the severity of their consequences on predictive accuracy,
is a question of crucial interest in any real-world application.
Prior literature suggests that posterior predictive checking
may be the most useful way to study the scale and nature of such issues
\citep{gelman13}.
It refers to testing whether the model predictions
are compatible with observed data.

\begin{figure}
\centering
\includegraphics[width=0.586\textwidth]{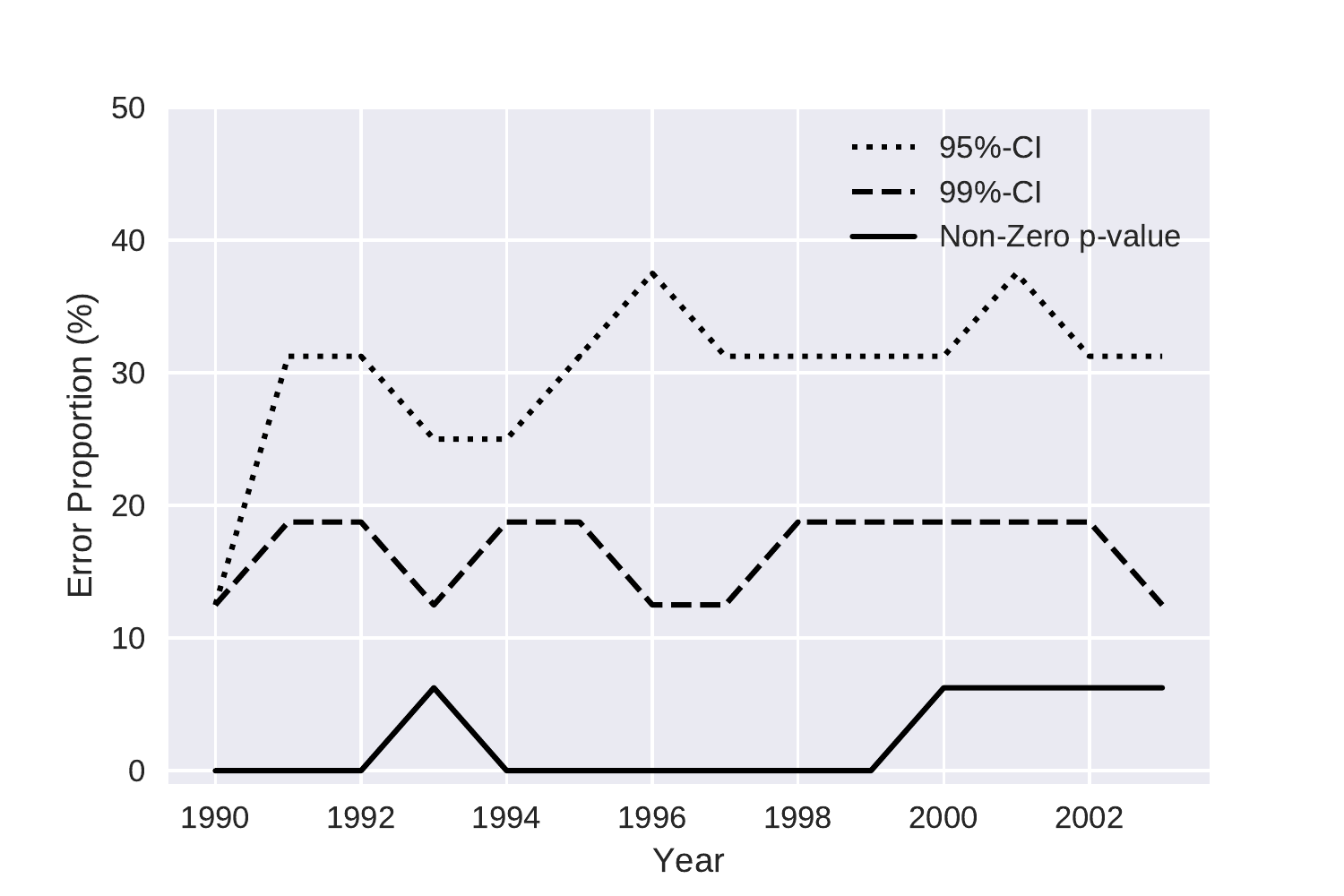}
\caption{BSC test of model validity on the German reunification data.
The graph visualizes findings from a relabeling exercise
where BSC was used to predict the comparison societies' outcomes.
The dotted and dashed lines indicate the proportion of outcomes
that fell outside their predicted 95\%-CI and 99\%-CI, respectively.
The solid line indicates the proportion of outcomes that
were estimated to have a posterior probability density of zero.}
\label{fig:deuppccheck}
\end{figure}

The relabeling exercise provides an obvious way
for running posterior predictive checks on BSC.
When the model is used to predict a comparison society post-reform trajectory,
the posterior predictive distribution should include the observed data in its spread.
Figure \ref{fig:deuppccheck} exhibits the results of this test.
The dotted and dashed lines indicate, for each year,
the share of comparison societies
for which the observed trajectory falls outside the 95\% and 99\%-CI, respectively.
The solid graph reflects the share of total prediction failures.
This refers to cases where the observed data is more extreme
than any single draw from the estimated posterior predictive distribution,
i.e. it receives the p-value of zero.

The share of predictions that lie within the 95\%-CI starts close to 90\%
in 1990, but soon falls to around two thirds and stabilizes at that level.
The wider 99\%-CI performs better, consistently capturing the observed data 80-90\% of the time.
Complete prediction failures are very infrequent,
with only a handful of years seeing even one single occurrence.
Perhaps surprisingly, none of the graphs demonstrates a clear upward trend over time.
The findings clearly demonstrate, first,
that the modeling assumptions are indeed violated.
Second, the importance of those violations is notable and
fairly constant over the studied 14-year timespan.

At the same time, the results are not altogether hopeless.
The confidence intervals include observed data most of the time,
even if less often than they should.
The 95\%-CI succeeds more than two thirds of the time
and the wider intervals perform better still.
This suggests that the linear factor model,
when accompanied by BSC's probabilistic structure,
is a useful even if imprecise model for GDP per capita growth in the OECD 1960-2003.

Note that this consistency check cannot be carried out for
SCM, GSC, ASCM, or any other method that uses up the relabeling findings
to calculate significance or confidence intervals.
They artificially inflate their confidence bounds
to include the comparison society observed data most of the time.
This may hide warning signs of assumption violations,
which is dangerous because both the point estimates and the confidence intervals
are valid only to the extent that the assumptions hold up.
Lack of validity checking makes analysis prone to overtly confident research conclusions.

\section{Application: California Tobacco Control Program}
\label{sec:california}

\subsection{Background}

The second canonical SCM paper examines the effect of a 1988 tobacco control reform
on California's cigarette consumption
\citep{abadie10}.
The reform, known as Proposition 99, introduced sin tax hikes and other anti-smoking measures.
The authors find that the reform's effect amounted to a 25\% fall in cigarette sales.
The study is famous for introducing the relabeling based significance test for SCM.
\cite{benmichael18} note the frequency at which the question has been re-analyzed.
We join in on this effort and study the same research question using BSC.

\subsection{Data}

The original authors' outcome variable
is the number of cigarette packs sold per capita (as per tax data).
As comparison societies they use a set of 38 other US states,
or all states that didn't introduce major tobacco controls of their own.
The time period covered is 1970-2000,
of which the years 1989-2000 form a treatment period for California.
For consistency, we use the same selection of data
and acquire it from a more recent edition of the publication used by the original authors
\citep{tobacco14}.
We ignore the other covariates included in the SCM analysis.

\subsection{Selecting the Number of Latent Factors}
\label{subsec:caliwaic}

Section \ref{sec:germany} fixed the number of latent factors at $L = 4$.
That choice of $L$ was computationally useful,
especially when running the heavy relabeling exercise,
but arbitrary from a modeling point of view.
In this application we propose a way of choosing $L$ through formal model selection.
Namely, the model can be run repeatedly with different values of $L$,
recording a measure of predictive performance at each round.
The choice of measure is not obvious,
but one robust option is the Watanabe-Akaike Information Criterion (WAIC).
WAIC is known to be asymptotically equivalent to measuring the model's
predictive accuracy with repeated cross-validation
\citep{watanabe10}.

The choice of $L$ in this section
begins with the \textit{a priori} assertion that $L \in \{ 3, 4, 5, 6, 7, 8 \}$.
The set is limited to be fairly small
because large values are computationally expensive.
Further, the variance of each latent factor's PCA-centered prior is proportional
to that of the PCA component,
so ever smaller for every additional included factor.
This urges caution against increasing the number of factors
endlessly even in the face of improving predictive accuracy,
so as not to introduce parameters with arbitrarily strongly informative priors.

The resulting WAIC values are collected in Table \ref{tab:caliwaic}.
The value is decreasing in the number of included factors.
Smaller WAIC indicates better predictive performance,
so we choose the model with the most factors: $L=8$.
The full prior specification for all other parameters
is again included in the Supplementary Materials.

\begin{table}
\begin{center}
\begin{tabular}{| c | c | c | c | c | c | c | c |}
\hline
$L$ & 3 & 4 & 5 & 6 & 7 & 8 \\ \hline
WAIC & 7308 & 6834 & 6616 & 6538 & 6450 & 6326\\ 
\hline
\end{tabular}
\end{center}
\caption{Model comparison findings for the number of latent factors.
The table records the Watanabe-Akaike Information Criterion (WAIC),
an asymptotic Bayesian model comparison statistic,
for six alternative BSC model specifications on the California dataset.
The specifications differ by the number of latent factors.
A smaller value indicates superior predictive performance.}
\label{tab:caliwaic}
\end{table}

\subsection{Findings}

\begin{figure}
\centerline{\includegraphics[width=0.586\textwidth]{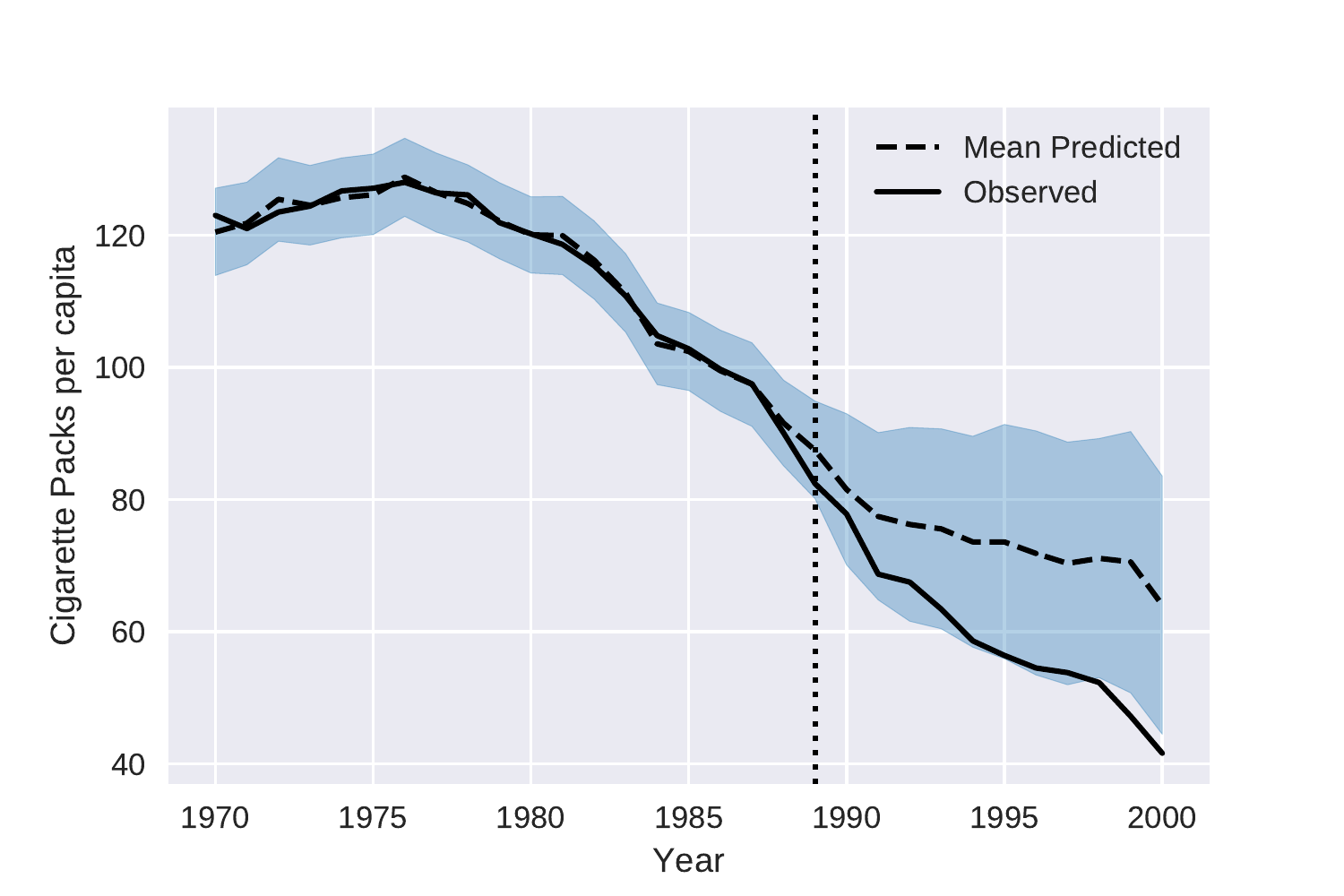}}
\caption{The BSC posterior predictive distribution of the counterfactual
cigarette consumption trajectory without the passing of Proposition 99.
The dashed trajectory and the shaded region represent the distribution's
mean and 95\%-confidence interval (CI).
The CI includes error re-sampling for the full time period.
Observed data is represented by the solid line.}
\label{fig:cabsccf}
\end{figure}

The core findings are captured in Figure \ref{fig:cabsccf} which
plots the mean predicted counterfactual trajectory and its 95\%-CI along with the observed data.
Like previous work,
the BSC model finds that the counterfactual trajectory falls slower than the observed smoking rate.
The two trajectories don't diverge notably for the first couple of years after the reform.
Starting around 1992, however, the gap begins to grow more substantial.
By 2000, the predicted rate is 64.0 packs per person,
or almost 22.4 packs (54\%), greater than the observed rate.

The BSC findings are quite similar to those of
\cite{abadie15}
when it comes to the scale of the treatment effect.
We find that the the reform reduced smoking over the 1989-2000 period
by 15.4 annual packs per person, or by 23\%.
The reported SCM estimate is slightly larger at approximately 25\%.
For further comparison,
\cite{benmichael18}
report some predicted counterfactual effects for the particular year 1997.
The predictions are 26 per capita for SCM
and 20 or 13 for two different Augmented SCM (ASCM) implementations.
The BSC mean estimate for 1997 is 16.5 packs,
so it falls well into the spread of the frequentist point estimates.

\begin{figure}
\centerline{\includegraphics[width=0.5\textwidth]{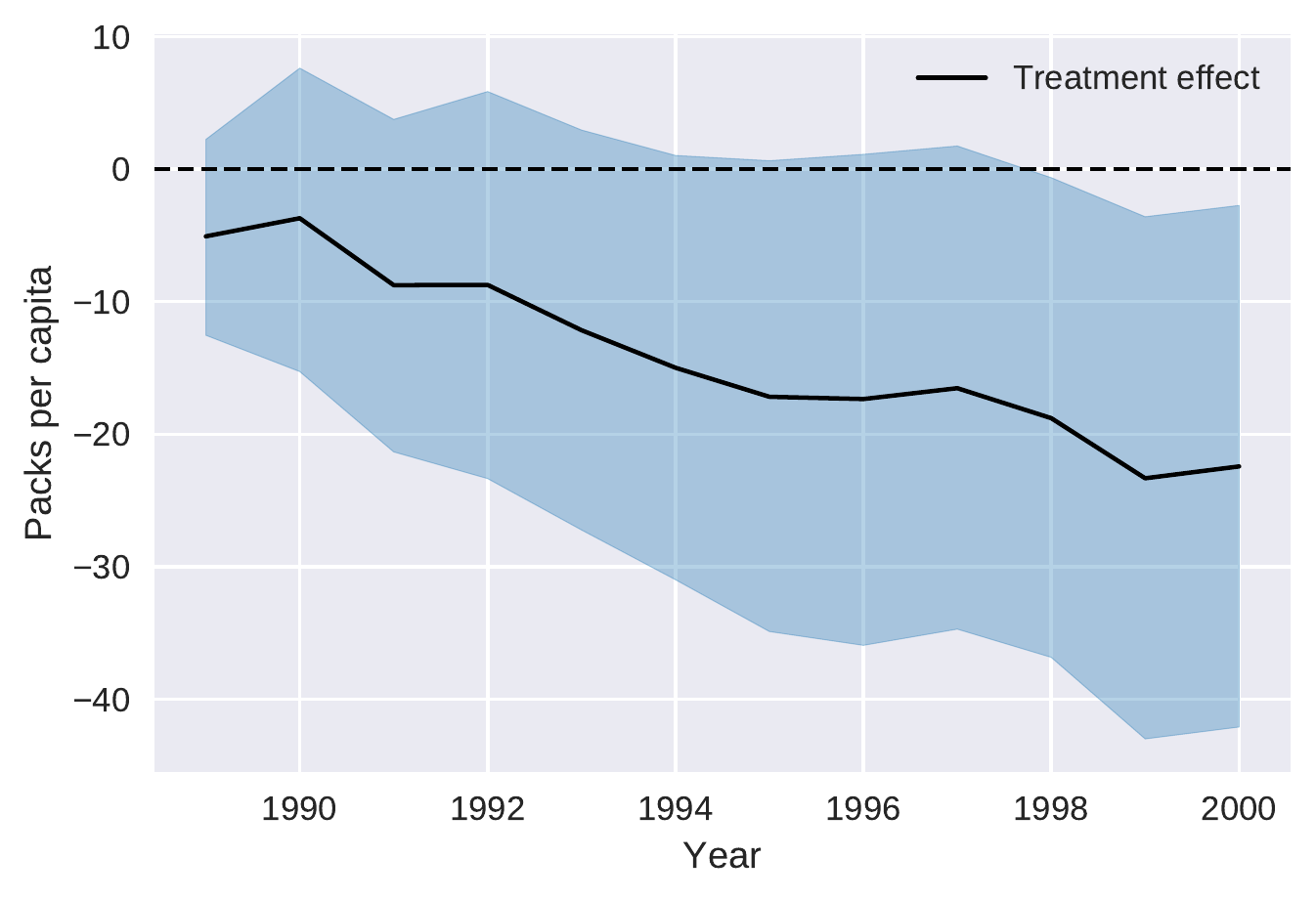}\includegraphics[width=0.5\textwidth]{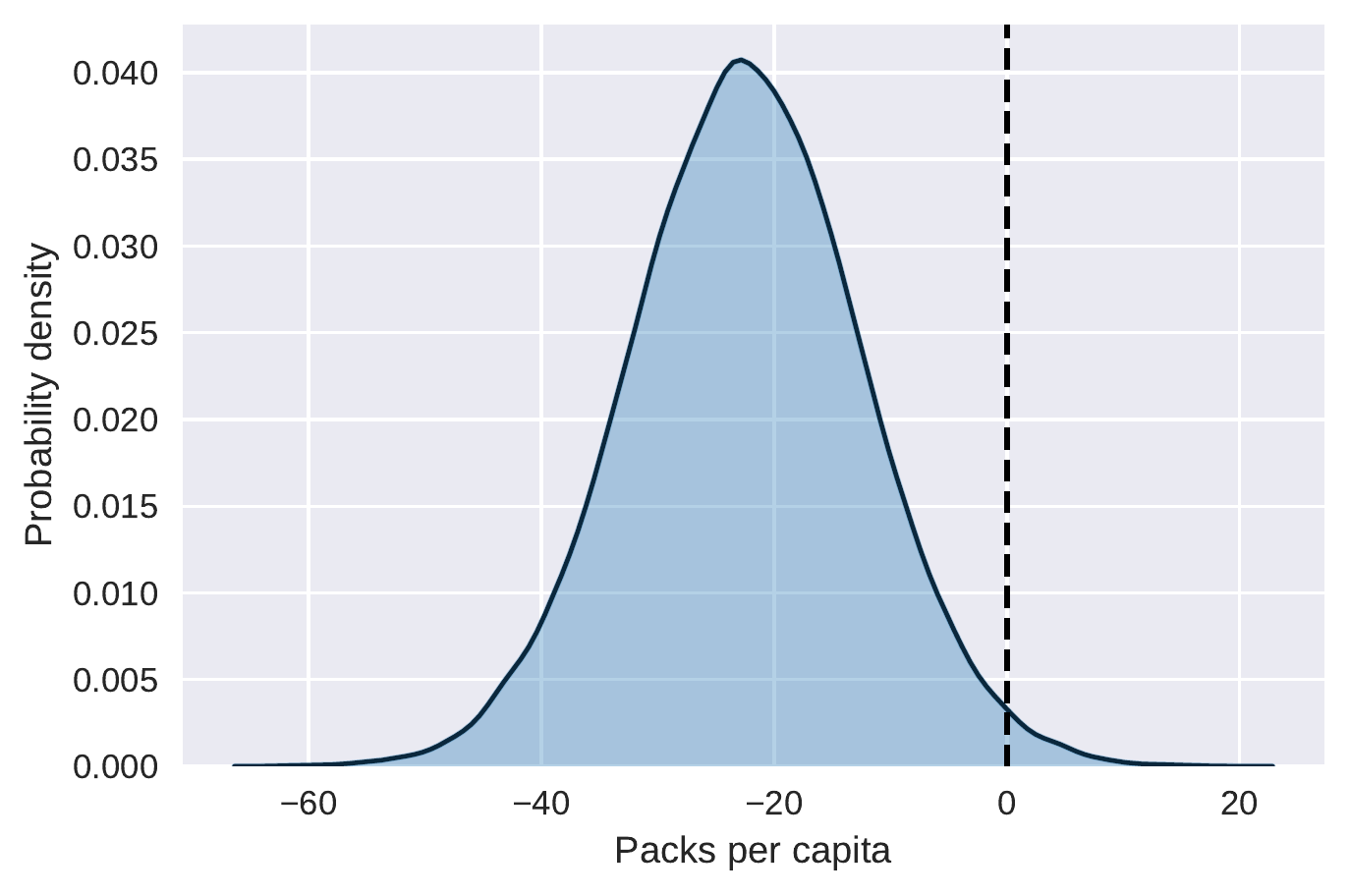}}
\caption{BSC posterior distribution of Proposition 99's treatment effect
on cigarette consumption in California.
Left panel plots the mean and 95\%-CI 
of the effect by year.
Right panel visualizes the full distribution
of the treatment effect at year 2000,
i. e. the aggregate effect over the full post-reform time period.}
\label{fig:caalpha}
\end{figure}

However, BSC's ability to quantify its uncertainty
reveals that the treatment effect's significance is dubious.
Note how the observed trajectory remains within the shaded 95\%-CI
throughout most of the post-reform time period.
Up to 1997, the model yields a probability of over 5\% that, even without the reform, 
tobacco consumption would have been as far from the prediction mean as the observed trajectory.
The treatment effect becomes significant in 1998-2000, but just barely so.
The full posterior distribution of the effect is illustrated in Figure \ref{fig:caalpha}.

These significance findings differ somewhat from those of previous studies.
\cite{abadie15} find stronger evidence,
with the treatment effect becoming significant as early as 1993.
\cite{benmichael18}, instead, conclude that the effect was insignificant throughout.
They exhibit a frequentist two-standard error interval
for each of their three tested frequentist SCM/ASCM specifications,
and show that a nonnegative effect falls well within each interval.
(It should be noted that the frequentist synthetic controls' sampling distributions are unknown,
so the two-standard error bound isn't guaranteed to correspond to a 95\%-CI.)

In conclusion, BSC's point estimates in the California tobacco control case
resemble those of previous researchers.
On statistical significance, however, BSC disagrees with its frequentist counterparts.
Unlike SCM, it finds the treatment effect insignificant
(in the Bayesian sense) for most of the post-reform time period.
Yet unlike ASCM, it deems the effect significant by the end year 2000, even if just barely so.
This disagreement between the various methods emphasizes
the importance of the choice of method for constructing synthetic controls.
In the face of such disagreement BSC is a very appealing choice,
especially because its credible intervals are valid on finite samples rather than asymptotically.

\section{Conclusion}
\label{sec:conclusion}

This paper's main contribution is the design for
the Bayesian Synthetic Control (BSC),
a novel statistical framework for the counterfactual estimation
problem thus far mostly addressed using the synthetic control method (SCM).
Like SCM and its frequentist extensions,
BSC is based on a linear factor model.
It has certain strengths over those preceding methods:
(1) it describes in full the associated prediction uncertainty,
including valid finite sample credible intervals;
(2) its Bayesian nature protects it from overfitting; and
(3) it enables the use of relabeling to check model validity.

We implement BSC on two previously studied research questions,
the German reunification and a California tobacco control program.
In the former case BSC yields similar findings as SCM
but outperforms it in a simple test of predictive accuracy.
This may be due to an overfitting issue in SCM.
This empirical section also illustrates BSC's unique ability
to use relabeling to assess model validity.
In the California case, BSC casts doubt on prior researchers' (mutually contradictory)
findings on statistical significance.
Together, the applications show that BSC is ready for implementation
in practical research settings.

The proposed framework continues to have notable limitations.
Importantly, unnecessarily restrictive PCA-based priors are used for latent factor trajectories.
This is done to aid computation in the Markov Chain Monte Carlo (MCMC) implementation.
Future research could examine whether a change of estimation strategy
(e.g. to variational inference or stochastic gradient MCMC)
would eliminate the need for this restriction.
Similarly, the proposed approach to the number of latent factors is not ideal:
Bayesian model averaging would likely be preferable to model selection using WAIC.
Both limitations are related to a third flaw,
characteristic of many modern Bayesian methods:
BSC is computationally much heavier than its frequentist counterparts.

Due to its Bayesian estimation goal,
that of describing the posterior (predictive) distribution,
BSC doesn't depend on the asymptotic qualities of any particular estimator.
Thus, its modeling assumptions can be seamlessly altered.
It is trivially easy to include two or more treated societies.
Missing data points in comparison societies could
be addressed with similar ease by marking them as stand-alone treated years,
a feature that markedly relaxes data availability restrictions.
Other extensions, like nonlinear factors,
nonconstant noise term variance,
and lagged outcomes or other time series behavior,
could also be included without rethinking the implementation strategy.

To our knowlegde, the present paper represents
the first explicit attempt to solve the synthetic control
counterfactual estimation problem for policy reforms in the Bayesian paradigm.
The model, though still subject to certain flaws,
is demonstrably ready for real world applications
and competitive in performance to existing frequentist tools.
This paper may then hopefully spur further research
into Bayesian solutions to the synthetic control problem
and other related topics in causal inference.

\bibliography{bscbiblio}
\bibliographystyle{apalike}

\end{document}


\maketitle
\thispagestyle{empty}

\setcounter{page}{1}
\setcounter{tocdepth}{2}

\begin{center}
Technical supplement.
\end{center}

\newpage

\tableofcontents

\newpage


\section{Bayesian Synthetic Control: Detailed Distributional Specification}

In this supplementary section we lay out in further detail the distributional assumptions
of the Bayesian Synthetic Control (BSC) probabilistic model.
First, we do so by component;
then, we state the set of equations that describes the whole system.

\subsection{Prior Distributions}

\subsubsection{White Noise}

Suppose that the random noise term $\varepsilon_{it}$
follows a normal distribution with mean zero
and an unknown standard deviation $\sigma$.
Assume that $\sigma$ is constant over societies and does not change over time.
Let the value of $\sigma$ have a half-Cauchy prior distribution
with the known scaling parameter $\gamma_\sigma$.

\subsubsection{Annual Fixed Effect}

Suppose that each annual fixed effect is drawn from a gaussian prior
with known mean and standard deviation.
Call the mean $\delta^{\mu}$ and the standard deviation $\delta^{sd}$.

\subsubsection{Intercept}

Set a hierarchical prior for the society fixed effects $\bm{K}$.
Namely, suppose that each society-specific intercept $\kappa_i$ is drawn
from a normal distribution with unknown mean $\kappa^{\mu}$
and standard deviation $\kappa^{sd}$.
Let the prior distribution of $\kappa^{\mu}$ be a gaussian with
the known mean $k_{\mu}$ and standard deviation $k_{sd}$.
For $\kappa^{sd}$, let the prior be half-Cauchy with the known scaling parameter $\gamma_\kappa$.

\subsubsection{Treatment Effect}
\label{subsubsec:effect}

Let each element $\alpha_{it}$ of the treatment effect matrix $\bm{\alpha}$
follow an independent gaussian prior with known mean $\alpha^{\mu}$ and standard deviation $\alpha^{sd}$.
It is extremely important to have this distribution be near-uniform
for all plausible treatment effect sizes,
i.e. to set $\alpha^{sd}$ equal to a very large number.
Otherwise the model will prefer a particular effect size
and adjust the $\bm{\beta}$ coefficients
such that the counterfactual trajectory is estimated accordingly.
This will make the model very sensitive to the treated trajectory,
a strongly undesirable feature for a model aimed to estimate the counterfactual trajectory.
It should pay little regard to the observations distorted by the intervention.

\subsubsection{Transformation Coefficients}

Set a hierarchical prior for the factor loadings $\bm{B}$.
Namely, let all country-specific coefficients $\beta_{im}$ for the $m$-th latent component
be drawn from a normal distribution with an unknown mean
$\beta_m^{\mu}$ and standard deviation $\beta_m^{sd}$.
For all $m = 1, \ldots, L$, let $\beta_m^{\mu}$ follow a gaussian prior
with the known mean $b_{\mu}$ and standard deviation $b_{sd}$,
and $\beta_m^{sd}$ to have a half-Cauchy prior with known scaling parameter $\gamma_{\beta}$.

\subsubsection{Latent Factors}

Let the value of the $m$-th latent factor in the year $t$
have a Gaussian prior centered around the known mean $p_{mt}$
with the fractor-specific standard deviation $r_m$.
For shorthand, denote by $\bm{P}$ the $T \times L$ matrix of these prior means,
by $\bm{r}$ the vector of length $L$ of standard deviations,
and by $\bm{R}$ the $T \times L$ matrix that results when $\bm{r}$
is stacked on top of itself $T$ times.
Ideally, we would have an uninformative prior such that
$p_{mt} = 0$ and $r_m = r_n$ for all $m, n, t$.
In the implementations of this paper,
however, we use an informative prior to aid computation for the MCMC sampler.

\subsection{Formal Model Specification}

The full BSC model can be precisely described in a single set of equations.

\subsubsection{Notation}

We denote vectors and matrices by boldface characters $(\bm{\kappa}, \bm{Y})$
and scalars by regular characters $(\kappa^\mu, \sigma)$.
Scalar names use a regular font $(\kappa^\mu, \sigma)$,
vectors are small case and in bold $(\bm{\kappa})$,
and bold upper case characters refer to matrices $(\bm{Y}, \bm{A})$.
All standalone vectors are treated as column vectors.
We refer to a vector of $X$ ones as $\bm{1_{\scaleto{[X]}{5pt}}}$ 
and to a matrix of $X \times Z$ ones as $\bm{1_{\scaleto{[X \times Z]}{5pt}}}$.

As usual, we use $\mathcal{N}$ for the normal distribution.
When the symbol appears with a dimension subscript, like this
$\mathcal{N}_{\scaleto{[A]}{5pt}}$ or this
$\mathcal{N}_{\scaleto{[A \times B]}{5pt}}$,
it refers to a vector or matrix of elementwise scalar gaussians.
Importantly, this doesn't refer to a multivariate or a matrix normal,
which input covariance matrices.
Instead, a vector or matrix of variances is supplied
and each element is drawn independently.
Implicitly, then, all covariances are assumed to be zero.
The same elementwise interpretation is taken to the squaring operator $[]^2$
when it modifies a vector or a matrix.

Here $\circ$ denotes the Hadamard product of two matrices of equal dimension.
Hadamard multiplies the input matrices' cells elementwise.
By $\otimes$ we refer to an outer product of two same-length vectors.
This is the exact opposite of the dot product:
if $\bm{x} \cdot \bm{z} = \bm{x}'\bm{z}$ yields a scalar,
then $\bm{x} \otimes \bm{z} = \bm{x}\bm{z}'$ yields a square matrix.
The outer product is used underneath exclusively for one purpose,
that of stacking a vector repeatedly on top of (or next to) itself to form a matrix like this:
$\bm{1_{\scaleto{[A]}{5pt}}} \otimes \bm{x}$ (or $\bm{x} \otimes \bm{1_{\scaleto{[A]}{5pt}}}$).

\subsubsection{Model}

\begin{align}
\label{eq:bsc_Y}
&\bm{Y} \sim \mathcal{N}_{\scaleto{[T \times J]}{5pt}} \left( \bm{M}, \bm{\Sigma}^2 \right),\\
\label{eq:bsc_sigmam}
&\bm{\Sigma} = \sigma \bm{1}_{\scaleto{[T \times J]}{5pt}},\\
\label{eq:bsc_sigma}
&\sigma \sim \text{Half-Cauchy} ( \gamma_{\sigma} ), \\
\label{eq:bsc_mu}
&\bm{M} = \bm{F} \bm{B}' + \bm{\Delta} + \bm{K} + \bm{A} \circ \bm{D},\\
\label{eq:bsc_L}
&\bm{F} \sim \mathcal{N}_{\scaleto{[T \times L]}{5pt}} \left( \bm{P}, \bm{R}^2 \right), \\
\label{eq:bsc_alpha}
&\bm{A} \sim \mathcal{N}_{\scaleto{[T \times J]}{5pt}} \left( \alpha^{\mu} \bm{1}_{\scaleto{[T \times J]}{5pt}}, \left(\alpha^{sd} \right)^2 \bm{1}_{\scaleto{[T \times J]}{5pt}}\right),\\
\label{eq:bsc_Delta}
&\bm{\Delta} = \bm{\delta} \otimes \bm{1}_{\scaleto{[J]}{5pt}},\\
\label{eq:bsc_delta}
&\bm{\delta} \sim \mathcal{N}_{\scaleto{[T]}{5pt}} \left( \delta^{\mu} \bm{1}_{\scaleto{[T]}{5pt}}, \left( \delta^{sd} \right)^2 \bm{1}_{\scaleto{[T]}{5pt}} \right), \\
\label{eq:bsc_K}
&\bm{K} = \bm{1}_{\scaleto{[T]}{5pt}} \otimes \bm{\kappa}, \\
\label{eq:bsc_kappa}
&\bm{\kappa} \sim \mathcal{N}_{\scaleto{[J]}{5pt}} \left( \kappa^{\mu} \bm{1}_{\scaleto{[J]}{5pt}}, \left( \kappa^{sd} \right)^2 \bm{1}_{\scaleto{[J]}{5pt}} \right), \\
\label{eq:bsc_kappamu}
&\kappa^{\mu} \sim \mathcal{N} \left( k_{\mu}, k_{sd}^2 \right) ,\\
\label{eq:bsc_kappasd}
&\kappa^{sd} \sim \text{Half-Cauchy} \left( \gamma_\kappa \right),\\
\label{eq:bsc_beta}
&\bm{B} \sim \mathcal{N}_{\scaleto{[J \times L]}{5pt}} \left( \bm{B^{\mu}}, \left( \bm{B^{sd}} \right)^2 \right) ,\\
\label{eq:bsc_betamu}
&\bm{B^{\mu}} = \bm{1}_{\scaleto{[J]}{5pt}} \otimes \bm{\beta^{\mu}}, \\
\label{eq:bsc_bmu}
&\bm{\beta^{\mu}} \sim \mathcal{N}_{\scaleto{[L]}{5pt}} \left( b_{\mu} \bm{1}_{\scaleto{[L]}{5pt}}, \left( b_{sd} \right)^2 \bm{1}_{\scaleto{[L]}{5pt}} \right), \\
\label{eq:bsc_betasd}
&\bm{B^{sd}} = \bm{1}_{\scaleto{[J]}{5pt}} \otimes \bm{\beta^{sd}}, \\
\label{eq:bsc_bsd}
&\bm{\beta^{sd}} \sim \text{Half-Cauchy}_{\scaleto{[L]}{5pt}} (\gamma_{b} \bm{1}_{\scaleto{[L]}{5pt}}).
\end{align}

Line \ref{eq:bsc_Y} above specifies the model's gaussian likelihood function,
and lines \ref{eq:bsc_mu} and \ref{eq:bsc_sigmam} specify its mean and standard deviation inputs.
The latter is not really a substantial equation,
but rather a description of filling up a matrix with the single scalar $\sigma$.
The prior of that variance term is expressed on line \ref{eq:bsc_sigma}.
Construction of the the mean term $\bm{M}$ is somewhat more convoluted.
The rest of the lines (\ref{eq:bsc_L} - \ref{eq:bsc_bsd})
are devoted for that task.

Line \ref{eq:bsc_L} specifies the prior of the latent factors,
and line \ref{eq:bsc_alpha} does the same for the treatment effect matrix.
Line \ref{eq:bsc_delta} defines the prior of the annual fixed effect vector, and
\ref{eq:bsc_Delta} contains notation for stacking that vector on top of itself
repeatedly to form the appropriately shaped matrix.
Line \ref{eq:bsc_kappa} defines the prior distribution of the society-specific intercept vector.
The (hyper)parameters of this distribution themselves have priors specified by \ref{eq:bsc_kappamu} and \ref{eq:bsc_kappasd}.
Again, line \ref{eq:bsc_K} simply stacks the intercept vector repeatedly into a matrix.
Lines \ref{eq:bsc_beta} - \ref{eq:bsc_bsd} work similarly:
\ref{eq:bsc_beta} states the prior distribution of the transition matrix of factor loadings,
\ref{eq:bsc_bmu} and \ref{eq:bsc_bsd} determine priors for the parameters of that distribution,
and \ref{eq:bsc_betamu} and \ref{eq:bsc_betasd} describe repeated stacking of vectors.

\newpage
\section{Parameter Specification for Empirical Applications}

A practical implementation of BSC in a concrete research setting
requires specifying the various 'known' prior parameters.
Here we record (and briefly discuss)
the values used in each of the two empirical applications of this paper.

\subsection{German Reunification}
\label{subsec:germany}

\subsubsection{White Noise}

The prior distribution of the white noise parameter $\sigma$
is half-Cauchy, which only takes one scaling hyperparameter $\gamma_\sigma$.
The half-Cauchy distribution has an infamously fat tail,
so the scaling parameter can be set with relatively little concern.
We opt for $\gamma_\sigma = 500$.
This corresponds to saying that we believe, \textit{a priori},
there to be $1/2$ probability that the country-year-specific
noise term is drawn from a normal with a standard deviation less than USD 500.

\subsubsection{Annual Fixed Effect}

The annual fixed effect has a constant mean $\delta^{\mu}$ for all years,
which we set equal to zero.
We suspect the annual fixed effect should never grow very large relative
to the overall size of the typical society.
However, the strength of that belief is moderated by the prospect that the annual fixed effect
interact one way or another with the latent factors.
To err on the side of ignorance, we thus set the prior standard deviation to $\delta^{sd} = 10,000$.

\subsubsection{Society Fixed Effect}

The intercept term indicates each country's mean income over the period 1960-2003.
Setting it up requires specifying three hyperparameters.
The unknown mean $\kappa^\mu$ is drawn from a gaussian with two hyperparameters:
mean $k_\mu$ and standard deviation $k_{sd}$.
The unknown standard deviation $\kappa^{sd}$ is drawn from a half-Cauchy
with the scaling parameter $\gamma_\kappa$.

A priori, we believe the average of the country mean incomes should be lower than
USD 30,000, a fairly typical Western per capita income around the turn of the millennium.
We also believe that it should be higher than USD 6,000, a definite lower bound for most developed countries.
To reflect this, we set $k_\mu = 18,000$ and $k_{sd} = 6,000$.
We indicate similar uncertainty over the variance of country means by setting
$\gamma_\kappa = 2,500$.

\subsubsection{Treatment Effect}

The year-country specific treatment effect $\alpha$ has a gaussian prior
with the known mean $\alpha^\mu$ and standard deviation $\alpha^{sd}$.
I set $\alpha^\mu = 0$ so as to not presuppose the sign of the effect.
Also recall from section \ref{subsubsec:effect} how vital it is
to set $\alpha^{sd}$ equal to some very large value.
In other words, it is important to make this prior very uninformative.
It should be almost flat over all even vaguely reasonable values.
Preferring to err on the side of unnecessarily flat,
we set $\alpha^{sd} = 30,000$.

\subsubsection{Factor Loadings}

The country-specific coefficient for each latent factor
is drawn from the same gaussian prior with unknown mean and standard deviation.
That mean itself has a gaussian hyperprior with mean $b_\mu$ and standard deviation $b_{sd}$,
while the standard deviation has a half-Cauchy hyperprior with the scaling parameter $\gamma_\beta$.
The scale of these loading coefficients is fundamentally linked
to the scale of the latent factor trajectories with which they are multiplied to generate observed data.
The factors themselves are centered around nonnormalized PCA components,
so unit variance is an appropriate scale for the coefficients.
Therefore, we fix: $b_\mu = 0$, $b_{sd} = 1$, and $\gamma_\beta = 1$.

\subsubsection{Latent Factors}

To ease computation for the Markov Chain Monte Carlo (MCMC) sampler,
each latent factor prior is pinned around a frequentist estimate for one PCA component.
This means that the latent factors $F$ have a heavily informative, data-driven prior.
We carry out the PCA analysis using a pre-existing implementation based on singular-value decomposition
from the \texttt{scipy} Python library.
We fix the number of latent components at $L = 4$.
The algorithm yields components centered around zero
with variance similar in magnitude to those of the vectors in the observed dataset.
The resulting components are stacked into a $T \times L$ matrix
and the mean of the latent factor prior $\bm{P}$ is set equal to that matrix.

It is less clear what the appropriate standard deviation $r_l$ is for each component.
The frequentist PCA components vary in variance,
so $r_l$ should be related to the variance of the underlying component.
We opt for direct proportionality where $r_l = \lambda \text{sd}^{pca}_{m}$.
The coefficient $\lambda$ should be as large (uninformative) as possible
while still ensuring that the factors are identifable.
In real terms we determined $\lambda$ heuristically by attempting to run the sampler with a few obvious guesses.
We concluded that $\lambda = 2$ is the largest integer for which the sampler consistently converges.
Thus, our prior for a latent variable trajectory allows its value to vary by two standard deviations
of the underlying PCA component's trajectory.

It should be noted that this specification of the $L$ prior
does not genuinely reflect prior uncertainty over its value,
but rather computational necessities.
Future research ought to explore computational estimation strategies other than conventional MCMC
in order to allow relaxing this strongly informative prior.

\subsection{California Tobacco Controls}

\subsubsection{White Noise}

Much of the US population was non-smokers throughout 1970-2000,
and most smokers adjust their smoking rate relatively little year-to-year.
Thus, we set a relatively conservative prior distribution on the white noise parameter $\sigma$.
Namely, we set $\gamma_\sigma = 10$ to reflect that we imagine the white noise term
to have a standard deviation less than ten packs per person with probability of one half.

\subsubsection{Annual Fixed Effect}

As in section \ref{subsec:germany},
we set the mean of the annual fixed effect term equal to zero.
To err on the side of ignorance, we nevertheless allow the prior standard deviation
to be quite large at $\delta^{sd} = 30$.

\subsubsection{Intercept}

We believe that the average of annual state smoking rates should be greater than zero
but less than 365, or a daily pack for each person.
To reflect this, we set $k_\mu = 180$, $k_{sd} = 90$, and $\gamma_\kappa = 90$.

\subsubsection{Treatment Effect}

As in section \ref{subsec:germany}, we opt for a vastly uninformative treatment effect prior:
$\alpha^\mu = 0$ and $\alpha^{sd} = 500$.

\subsubsection{Factor Loadings and Latent Factors}

We set transformation coefficient priors
and latent variable priors exactly as in section \ref{subsec:germany}.
Namely, we have $b_\mu = 0$, $b_{sd} = 1$, and $\gamma_\beta = 1$ for the coefficients,
and we set the factor prior means equal to PCA estimates
and factor prior standard deviations equal to twice the standard deviations of the associated PCA components.

\newpage

\section{Computation}

The particular MCMC algorithm used in both empirical applications was the No-U-Turns Sampler (NUTS).
NUTS is a modern extension to the famous Hamiltonian Monte Carlo (HMC) sampler;
it carries out adaptive parameter specification for the underlying HMC step function
\citep{hoffman14}.
We used a pre-existing NUTS implementation from the open-source \texttt{pymc3} Python library.
We ran the sampler with a target acceptance rate of 0.9
and a maximum tree depth of 12
in two parallel Markov chains,
each of which was given 5,000 tune-in steps.
Each chain was then run for 25,000 sampling steps,
for a total sample size of 50,000 per model.
The results were checked for sampler covergence
using the Gelman-Rubin diagnostic \citep{gelman92}
and the lack of NUTS sampler divergences.

The \texttt{pymc3} library was used for most aspects
of specifying and estimating the probabilistic model.
The library \texttt{sklearn} was used to carry out PCA analysis
through the library's singular value decomposition-based PCA algorithm.
Other libraries on which the implementation code depends include
\texttt{numpy}, \texttt{pandas}, \texttt{theano}, \texttt{matplotlib}, and \texttt{seaborn}.

Running the implementation code was computationally quite heavy.
It was done on an Amazon Web Services (AWS)
virtual Ubuntu server equipped with substantial processing power.
Each run of the baseline BSC implementation in the empirical sections took 8-12 hours.
This was quite consequential for the relabeling exercise in the German application
and the WAIC comparison in the California application,
because they involved running the model repeatedly 16 times
(once for each comparison society)
and 6 times (once for each possible value of $L$), respectively.
The total computation time was therefore just over one week.

\newpage
\bibliography{bscbiblio}
\bibliographystyle{apalike}
\addcontentsline{toc}{section}{References}